\documentclass[12pt,letterpaper]{article}
\usepackage{amsfonts,amssymb,amsmath,bbm,euscript,mathrsfs,calligra}
\usepackage{bm,mathrsfs}
\usepackage[usenames]{color}
\usepackage{color,soul}
\usepackage[utf8]{inputenc}
\usepackage[T1]{fontenc}
\usepackage{textcomp}
\usepackage{graphicx}
\usepackage[tight,sf]{subfigure}
\usepackage{listings}
\usepackage{hyperref}
\usepackage{bookmark}

\DeclareMathAlphabet{\mathcalligra}{T1}{calligra}{m}{n}
\DeclareFontShape{T1}{calligra}{m}{n}{<->s*[2.2]callig15}{}

\newcommand{\sech}{\mathrm{sech} \,}

\newcommand{\arcsinh}{\mathrm{arcsinh} \,}

\title{\Huge{Environmental bias and elastic curves on surfaces}}

\author{\Large
Jemal Guven,${}^1$\footnote{\href{mailto:jemal@nucleares.unam.mx}{jemal@nucleares.unam.mx}}
Dulce Mar\'ia Valencia${}^2$\footnote{\href{mailto:dvalencia@fis.cinvestav.mx}{dvalencia@fis.cinvestav.mx}}
and Pablo V\'azquez-Montejo${}^3$\footnote{\href{mailto:pablov@andrew.cmu.edu}{pablov@andrew.cmu.edu}}
}
\date{}

\begin{document}

\maketitle
\begin{center}
{\it
$^1$ Instituto de Ciencias Nucleares, Universidad Nacional Autónoma de México\\
 Apdo. Postal 70-543, 04510 M\'exico D.F., MEXICO\\
$^2$ Departamento de Física, Cinvestav,\\ Instituto Politécnico Nacional 2508,
07360, M\'exico D.F., MEXICO\\
$^3$ Department of Physics, Carnegie Mellon University,\\
5000 Forbes Avenue, Pittsburgh, PA 15213, USA}
\end{center}

\begin{abstract}
The behavior of an elastic curve bound to a surface will reflect the geometry of its environment. This may occur in an obvious way: the curve may deform freely along directions tangent to the surface, but not along the surface normal. However, even if the energy itself is symmetric in the curve's geodesic and normal curvatures, which  control these modes, very distinct roles are played by the two. If the elastic curve binds preferentially on one side, or is itself assembled on the surface, not only would one expect the bending moduli associated with the two modes to differ, binding along specific directions, reflected in spontaneous values of these curvatures, may be favored. The shape equations describing the equilibrium states of a surface curve described by an elastic energy accommodating environmental factors will be identified by adapting the method of Lagrange multipliers to the Darboux frame associated with the curve. The forces transmitted to the surface along the surface normal will be determined.
Features associated with a number of different energies, both of physical relevance and of mathematical interest, are described. The conservation laws associated with trajectories on surface geometries exhibiting continuous symmetries are also examined.
\end{abstract}

\section{Introduction}

Linear structures are frequently encountered on surfaces. A  familiar example is provided by the interface separating two phases of a fluid membrane \cite{Helfrich}, described by an energy proportional to its length \cite{Lipowsky}.  Resistance to bending becomes relevant in the modeling of a semi-flexible linear polymer bound to a surface  along  its length or a linear protein complex that self-assembles on the surface.   Even if we confine our attention to its purely geometrical degrees of freedom--typically those that turn out to be the most relevant when we zoom  out to mesoscopic scales \cite{NPW, Markus}--the identification of the appropriate energy, nevermind attending to the dynamical behavior it predicts, is not obvious.  For the environment will play a role and  there is no single--physically relevant--analogue of the Euler elastic bending energy of a space curve,  quadratic in the three-dimensional Frenet curvature $\kappa$ along the curve, \cite{Singer, Kamien}
\begin{equation} \label{HBF}
 H_B =\frac{1}{2}\,  \int ds \, \kappa^2\,,
\end{equation}
where $s$ is arc-length. The mathematically obvious generalization, quadratic in the geodesic curvature $\kappa_g$--the analogue of $\kappa$ for a surface curve,
\begin{equation} \label{HG}
 H_g =\frac{1}{2}\,  \int ds \, \kappa_g^2\,,
\end{equation}
while interesting in its own right, is sensitive only to the metrical properties of the surface: it quantifies the bending energy of deformations tangential to the surface but fails to respond to how the surface itself bends in space; as such it also fails to capture the physics.
\vskip1pc \noindent
The Euler elastic energy (\ref{HBF}) of a surface bound curve  does better in this respect, capturing the environmental bias associated with contact that breaks the symmetry between the two bending modes implicit in the energy. To see this, recall that  the squared Frenet curvature can be decomposed into a sum of  squares $\kappa^2=\kappa_g^2 + \kappa_n^2$ \cite{DoCarmo}: the normal curvature $\kappa_n$, registers how the surface itself bends along the tangent to the curve,  thus picking up the curvature deficit not captured by $\kappa_g$.  Unlike $\kappa_g$, it is also fixed at any point once the tangent direction is established. In fact,  its maximum and minimum are set  by the surface principal curvatures; as such, and unlike geodesic curvature, it is bounded along a curve if the surface is smooth.
\vskip1pc \noindent
On the appropriate length scale, the constrained three-dimensional bending energy (\ref{HBF}) does provide  a reasonable physical description of a thin wire--of circular cross-section--bounding a soap film, or the elastic properties of a semi-flexible polymer obliged to negotiate an obstacle, a consequence of its confinement within the surface itself or its non-specific binding to it \cite{Mann, MannNick, GiomiMaha, Spakowitz, GuvVaz}.  Intrinsic and extrinsic bending energies, characterized by $\kappa_g^2$ and $\kappa_n^2$ respectively, are weighed equally.  Despite the striking difference in the natures of geodesic and normal curvatures they are treated symmetrically.
\vskip1pc \noindent
If a Euler elastic curve is often a poor caricature of a one-dimensional elastic object freely inhabiting three-dimensional space (DNA say), it fares worse as a description of a one-dimensional object inhabiting a surface. For implicit in the treatment of the two curvatures that places them on an equal footing is the assumption that the one-dimensional object possesses not only an existence but also material properties (albeit only its rigidity) that are independent of the surface constraining its movement. But in the case of protein complexes, assembling  into extended one-dimensional structures on a fluid membrane or condensing along its  boundary, the surface is clearly not a passive substrate. The interaction of the structure with the membrane will  be reflected in the elastic properties;  at a minimum, its rigidities tangential and normal the surface will differ, which establishes a distinction between how they bend intrinsically and extrinsically on the surface \cite{KamienNelson}.
\vskip1pc \noindent
If the substrate is a round sphere, all directions are equivalent and the normal curvature is constant. In this case,  the change  amounts to a simple recalibration of the energy.\footnote{ But, even if the substrate is metrically spherical,  its geodesics may be non-trivial and its normal curvature need not be constant.}  In general, however,  there will be physically observable consequences.  Consider for example an elastic curve on a minimal surface.  Locally, such surfaces minimize area; under appropriate conditions, they also minimize the surface bending energy of a symmetric fluid membrane; as such, they feature frequently in the morphology of biological membranes (see, for example, \cite{Hyde}). Because  its mean curvature vanishes,  the surface assumes a symmetric saddle shape almost everywhere. A large normal bending modulus would tend to favor low values of $\kappa_n$ so that the equilibrium tilts to align along the ``flat'' asymptotic directions, where $\kappa_n=0$; a large geodesic bending
modulus, on the other hand, would promote alignment along geodesics.
\vskip1pc \noindent
Any surface curve carries a {\it Darboux frame}.  If one of its two normals is chosen to lie along the surface normal,  its second co-normal is tangential.  The normal and geodesic curvatures measure how fast the tangent vector rotates into the surface normal and into the co-normal respectively as the curve is followed. This frame also introduces a geometrical notion of twist: the geodesic torsion $\tau_g$ (the rate of rotation of the surface normal into the co-normal); as such, it captures the deviation of the tangent  vector from the surface principal directions.\footnote {These are the directions on the surface along which the normal curvature is a maximum or minimum.} On a surface assembled curve, it may be appropriate to admit a dependence on this torsion.  To justify this claim note that the normal curvature and the geodesic torsion are not independent.  On a minimal surface, their squares sum  to give the manifestly negative Gaussian curvature ${\cal K}_G$ by\footnote{The Gaussian curvature is given
by the product of the two surface principal curvatures $C_1$ and $C_2$, ${\cal K}_G=C_1 \,C_2$, so is a surface property; it does not depend on the curve followed.}
\begin{equation} \label{KGkntg}
{\cal K}_G = -(\kappa_n^2+\tau_g^2)\,.
\end{equation}
Like $\kappa_n$, $\tau_g$ is bounded by the surface curvature.
\vskip1pc \noindent
On an isotropic fluid membrane there is a single natural spontaneous curvature \cite{Helfrich}; likewise elastic planar or space curves can have  a preferred curvature  \cite{SolisOlvera, CallanBrunAudoly, ArroyoGaray, Miller}. However, the interaction with the substrate may also involve the introduction of effective spontaneous curvatures which are different along tangential and normal directions. For instance, rod-like proteins may assume a nematic order \cite{Nelson, FrankKardar, KulicDeserno, Ipsen}. If these proteins condense--for one reason or another--into one-dimensional extended structures,  they may possess not only distinct bending moduli, as in a rod \cite{LandauLifshitz, LangerSinger, AudolyPomeau}, but also an anisotropic spontaneous curvature if the sub-units are curved along their length and bind preferentially along one side.\footnote{Unlike a space curve, along which a spontaneous Frenet curvature is ill-defined, there is no such ambiguity along surface-bound curves due to its inheritance
of the Darboux frame adapted to its surface environment.} A spontaneous geodesic torsion would promote deviation from the principal directions and, as a consequence the formation of spiral structures on surfaces of negative Gaussian curvature.
\vskip1pc \noindent
A natural generalization of the bending energy (\ref{HBF}) for a surface bound curve is
\begin{equation}\label{HBD}
H =  \frac{1}{2}\, \int ds \left((\kappa_g - C_g)^2 + \mu (\kappa_n - C_n)^2 + \nu (\tau_g -  C_0 )^2 \right)\,,
\end{equation}
involving relative normal and torsional rigidity moduli $\mu$ and $\nu$,  as well as  constant spontaneous geodesic and normal curvatures, $C_g$ and $C_n$, and torsion, $C_0$.  This, of course, is not the most general quadratic since it does not accommodate off-diagonal elements.
\vskip1pc \noindent
The Euler-Lagrange (EL) equation describing the equilibrium states of a curve  with an energy of the form (\ref{HBD}) are  presented in section 2.  In this outing, the background geometry will be assumed fixed,  but is otherwise arbitrary.  A direct approach  would be
to adapt the method developed by  Nickerson and Manning in the context of Euler-Elastic space curves negotiating an obstacle \cite{Mann, MannNick}. However,  we will examine the problem by extending an  approach, introduced by two of the authors to examine confined Euler elastic curves.  In this approach  track is kept of the chain of connections between the curve, the surface, and the Euclidean background. This permits the breakdown of Euclidean symmetry to be correlated with the source of tension in the curve:  (minus) the normal forces transmitted to the surface \cite{GuvVaz}.   For an energy of the form (\ref{HBD}), it would appear to be meaningless to talk about the breaking  of Euclidean symmetry,  because such an energy  cannot be defined without reference to  the surface environment.  Yet it is still useful to think in terms of the tension along the curve. In distinction to bound  Euler elastica, the normal force transmitted to the confining surface is not the only source of tension in the curve.  In
general, it will also be subjected to geometrical tangential forces along its length. Significantly,  this new tangential source of tension vanishes only when the functional dependence of the energy can be cast in terms of the Frenet curvature.
\vskip1pc \noindent
We will identify the boundary conditions that are physically relevant on open curves. We follow this with an examination of several cases of special  interest, among them energies that have been been treated in the recent literature.  Various possibilities of mathematical interest are also suggested by our framework, among them generalizations of the elastic energies on Riemannian manifolds examined by Langer and Singer \cite{LangSing} depending only on the surface intrinsic geometry. One such is to take  $\mu=0$ and $\nu=0$ in Eq. (\ref{HBD}), and add a potential depending on the local Gaussian curvature. Whereas this intrinsic dependence is reflected in the Euler-Lagrange equations, the forces transmitted to the surface do depend explicitly on the embedding in space.  We also examine a natural generalization of the Euler energy, with $\mu=1$, $\nu=1$ and vanishing spontaneous curvatures in Eq. (\ref{HBD}), so that it is symmetric not only with respect to the interchange of  the curvatures but also the
interchange of curvature with torsion.
\vskip1pc \noindent
We next examine background geometries preserving some subgroup of the Euclidean group. There will now be a conserved Noether current associated with each unbroken symmetry, which provides a first  integral of the shape equation. In particular, we examine curves on axially-symmetric surfaces where the conserved current is a torque. We also examine elastic curves on a helicoid,  exploiting the glide rotational symmetry of this geometry to identify the corresponding conserved quantity.
\vskip1pc \noindent
Various relevant identities are collected in a set of appendices.  Here the less indulgent reader will also find a Hamiltonian analysis adapted to the symmetry,  which is also potentially useful  for numerical solution of the Euler-Lagrange equations.

\section{Equilibrium states }

\begin{figure}[htb]
\begin{center}
\includegraphics[scale=0.75]{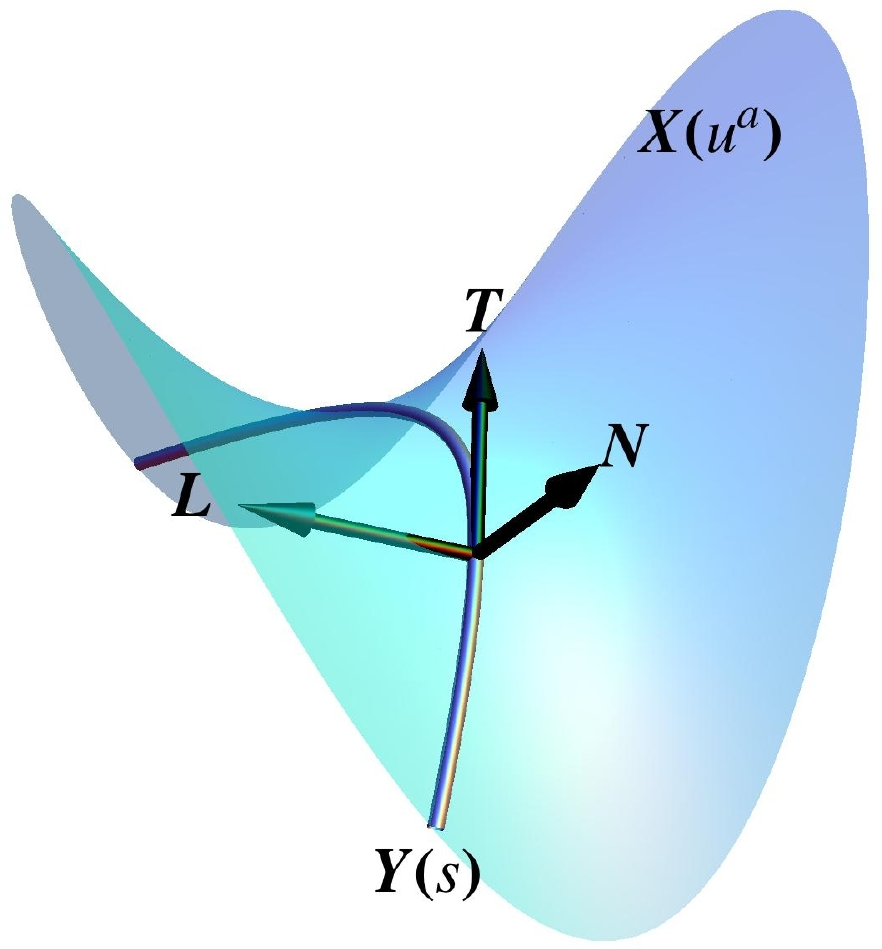}
\end{center}
\caption{\small{Curve ${\bf Y}(s)$ on a surface ${\bf X}(u^a)$ and its associated Darboux frame $\{{\bf T},{\bf L},{\bf N}\}$.}}
\label{Fig1}
\end{figure}

Let the surface $\Sigma$ be described parametrically, $\Sigma: (u^1,u^2) \mapsto \mathbf{X} (u^1,u^2)$, and let $\mathbf{e}_a = \partial_a \mathbf{X}$, $a=1,2$ be the two tangent vectors to this surface adapted to the parametrization, and $\mathbf{n}$ the unit vector normal to the surface (pointing outwards if it is closed). The surface curve $\Gamma$ is parametrized by arc-length, $\Gamma: s \mapsto (U^1(s),U^2(s))$, which can be identified as a space curve under composition of maps, $s \to \mathbf{Y}(s) = \mathbf{X}(U^1(s),U^2(s))$.
\vskip1pc \noindent
Let $\mathbf{T}= \mathbf{Y}'$ be the unit tangent vector to $\Gamma$ and $\mathbf{N}=\mathbf{n}(U(s))$ the unit vector normal to the surface along the curve.\footnote{Here, and elsewhere, prime represents derivation with respect to $s$.}  The associated Darboux frame is then defined by $\{\mathbf{T},\mathbf{L}, \mathbf{N}$, where $\mathbf{L}=\mathbf{N}\times \mathbf{T}\}$ (see Fig. \ref{Fig1}).
The structure equations (analogues of the Frenet-Serret equations) describing the rotation of this frame as the curve is followed are given by \cite{DoCarmo}
\begin{equation} \label{Darbouxequations}
\mathbf{T}' = \kappa_g \mathbf{L} - \kappa_n \mathbf{N}\,,\quad
\mathbf{L}' = -\kappa_g \mathbf{T} + \tau_g \mathbf {N} \,,\quad
\mathbf{N}' = \kappa_n \mathbf{T} - \tau_g \mathbf{L} \,.
\end{equation}
The geodesic curvature, $\kappa_g$, involves acceleration and thus two derivatives along the curve, whereas the normal curvature, $\kappa_n$, and the geodesic torsion, $\tau_g$ depend only on the tangent vector, and thus involve a single derivative. The geodesic torsion is significantly different from its Frenet counterpart, which involves three derivatives, in this respect. The two are related as follows: $\tau =\tau_g -  \theta'$, where $\theta$ is the angle rotating one frame into the other \footnote{The Frenet frame $\{\mathbf{T}, \mathbf{\cal N}, \mathbf{\cal B}\}$ is given in terms of the Darboux frame  by $\mathbf{\mathcal N} = \cos \theta \mathbf{L} - \sin \theta \mathbf{N}$ and $\mathbf{\mathcal B} = \sin \theta \mathbf{L} + \cos \theta \mathbf{N}$.}; the missing order in derivatives is  captured by a derivative of $\theta$.
\vskip1pc \noindent
More explicitly, the geodesic curvature can be cast in the intrinsic form, $\kappa_g = \mathbf{T}' \cdot \mathbf{L} = l^a t^b\nabla_b t_a$, where $\nabla_a$ is the covariant derivative compatible with the metric induced on the surface, $g_{ab}=\mathbf{e}_a\cdot\mathbf{e}_b$; $\kappa_n$ and $\tau_g$ involve the surface extrinsic curvature $K_{ab}=\mathbf{e}_a\cdot\partial_b \mathbf{n}$ through the identifications\footnote{We can expand $\mathbf{T}$ with respect to the basis of vectors adapted to the surface parametrization, $\{\mathbf{e}_a, a=1,2\}$, as follows $\mathbf{T}= t^a \mathbf{e}_a$, where $t^a=U^a{}'$. Similarly, one can expand $\mathbf{L}=l^a {\bf e}_a$.}
\begin{equation} \label{kaptaudef}
\kappa_n = \mathbf{N}' \cdot \mathbf{T}= t^a t^b K_{ab}\,\quad \tau_g = \mathbf{L}' \cdot \mathbf{N} = - l^a t^b K_{ab}\,.
\end{equation}
We now identify  the equations describing the equilibrium states of a surface curve with an energy given by\footnote{Our description of the variational principle is also valid if ${\cal H}$ depends on derivatives of the curvatures and torsion.}
 \begin{equation} \label{Hcurve}
H = \int ds\, \left( {\cal H}(\kappa_g, \kappa_n, \tau_g) + V\right) \,,
\end{equation}
where, for the moment, ${\cal H}$ is an arbitrary function of its three arguments, and $V$ is some potential that depends on the position upon the surface but not on the tangent vectors. These dependence could be through the curvatures of the surface. The important point is that the Hamiltonian (\ref{Hcurve}) depends only on the geometric degrees of freedom associated with the curve on the surface. To begin with we will not make any assumptions concerning the symmetry of the surface. Because we do not need to.

\subsection{The Euler-Lagrange equation and the normal force: results}

To examine the response of the energy (\ref{Hcurve}) to a deformation of the curve we extend the simpler framework developed by two of the authors to study the confinement of elastic curves described by the energy (\ref{HBF}), quadratic in the Frenet curvature \cite{GuvVaz}. The Frenet frame is, of course, a poor choice if $\kappa_g$ and $\kappa_n$ are treated asymmetrically, never mind contemplating an explicit dependence on $\tau_g$; nor is it sufficient to simply tweak the framework slightly to accommodate a Darboux frame. The interested reader will find the complete details of the derivation in Sect. \ref{Appderivation}.
\vskip1pc \noindent
The tension along the curve is given by the vector $\mathbf{F} = F^\mathbf{T} \mathbf{T} + F^\mathbf{L} \mathbf{L} + F^\mathbf{N} \mathbf{N}$, where the components are given by
\begin{subequations} \label{FTNL}
\begin{eqnarray}
F^\mathbf{T} &=& \kappa_g {\cal H}_g + \kappa_n {\cal H}_n + \tau_g {\cal T}_g  - {\cal H} - V - c\,,\\
F^\mathbf{L} &=& {\cal H}_g' + \tau_g  {\cal H}_n - \kappa_n {\cal T}_g\,,\\
F^\mathbf{N} &=& -{\cal H}'_n - \frac{\tau_g}{\kappa_n} \left({\cal T}_g' - \kappa_g {\cal H}_n \right)
 - \kappa_g {\cal T}_g + \frac{V'}{\kappa_n}\,;
\end{eqnarray}
\end{subequations}
the parameter $c$ is a constant associated with the fixed length of the surface curve, and\footnote{Partial derivatives  are replaced by their functional counterparts if a dependence on derivatives  with respect to arc-length is contemplated.}
\begin{equation} \label{Lambdatnnt}
{\cal H}_g  = \frac{\partial {\cal H}}{\partial \kappa_g} \,, \quad
{\cal H}_n = \frac{\partial {\cal H}}{\partial \kappa_n}\,,\quad
{\cal T}_g  = \frac{\partial {\cal H}}{\partial \tau_g}\,.
\end{equation}
In equilibrium the tension is not conserved, $\mathbf {F}'=-\bm \lambda$, where $\bm \lambda$ is an external force  normal to the curve due to its shaping by the surface geometry,
$\bm \lambda = \lambda^\mathbf{L} \mathbf{L} +  \lambda^\mathbf{N} \mathbf{N}$.  Its component tangent to the surface given by
\begin{equation} \label{eq:lamL0}
\lambda^\mathbf{L} = \frac{ {\cal K}_G}{\kappa_n} \, ({\cal T}_g' + \kappa_n {\cal H}_g - \kappa_g {\cal H}_n ) + \frac{\tau_g V' }{\kappa_n} + \nabla_\mathbf{L} V \,,
\end{equation}
where $\nabla_\mathbf{L} V = l^a \partial_a V$ is the directional derivative of $V$ along $\mathbf{L}$.  This tangential source of tension vanishes whenever the energy depends only on the Frenet curvature and in particular, for the Euler elastic bending energy (\ref{HBF}).  The  EL equation,  describing equilibrium states, is given by $\varepsilon_\mathbf{L} :=
\mathbf{L}\cdot \mathbf{F}' + \lambda^\mathbf{L}=0$. Collecting terms, one can express
\begin{align} \label{EL}
\varepsilon_\mathbf{L} &={\cal H}''_g + \frac{(\tau_g {\cal H}^2_n)'}{{\cal H}_n} +(K - 2 \kappa_n) ({\cal T}_g'- \kappa_g {\cal H}_n) + \left({\cal K}_G + \kappa^2_g \right) {\cal H}_g \nonumber \\
& -(\kappa'_n - 2 \kappa_g \tau_g) {\cal T}_g - \kappa_g ({\cal H} + V +c)+ \nabla_\mathbf{L} V =0\,.
\end{align}
Here $K$ is twice the local mean curvature.\footnote{It is given by the sum of the two principal curvatures, $K=C_1 + C_2$.} The magnitude of the normal force is given by $\lambda^\mathbf{N} =  - \mathbf{N}\cdot \mathbf {F}' $, or explicitly
\begin{align} \label{NF}
 \lambda^\mathbf{N} &={\cal H}''_n +\left( \frac{\tau_g}{\kappa_n} \left({\cal T}_g' - \kappa_g {\cal H}_n \right)
 + \kappa_g {\cal T}_g - \frac{V'}{\kappa_n} \right)' - \tau_g {\cal H}'_g \nonumber \\
 & + \kappa_g \kappa_n {\cal H}_g + (\kappa^2_n -\tau^2_g) {\cal H}_n + 2 \kappa_n \tau_g {\cal T}_g - \kappa_n ( {\cal H} + V + c) \,.
\end{align}
It also is completely determined by the geometry.

\subsection{Derivation} \label{Appderivation}

Three Lagrange multipliers (given as the components of a vector $\mathbf{F}$) are introduced to identify the tangent vector $\mathbf{T}$ with $\mathbf{Y}'$, another six ($\lambda_{IJ}$) to identify $\{\mathbf{T}, \mathbf{L}, \mathbf{N}\}$ as an orthonormal frame. Three further multipliers ${\cal H}_g, {\cal H}_n$ and ${\cal T}_g$ define the curvatures and torsion in terms of this frame.\footnote{To lighten the notational burden we use the same symbol for the multiplier as we do for the solutions of the unconstrained EL equations for $\kappa_g$, $\kappa_n$ and $\tau_g$.} A vector-valued Lagrange multiplier $\bm{\lambda}$ identifies the space curve with a curve on the surface, $\mathbf{Y}(s)= \mathbf{X}(U^1(s),U^2(s))$,  just as it did in \cite{GuvVaz}  for a constrained Euler elastic curve;  thus far this closely follows the derivation of the shape equation there.
\vskip1pc \noindent
What distinguishes the variational principle adapted to the Darboux frame from its Frenet counterpart is the need to  introduce another vector-valued multiplier, $\mathbf{\Lambda}$, to identify the frame field $\mathbf{N}(s)$ with the surface normal $\mathbf{n}(U(s))$.\footnote{An alternative constraint, equivalent to the latter, is provided by the term $\int ds\, \, \Lambda_\perp^a \mathbf{N}\cdot \mathbf{e}_a (U^a) ds$, where the tangent vectors $\mathbf{e}_a$ are treated as functionals of $U^a$: $\mathbf{e}_a(U^a) = \frac{\partial \mathbf{X}(u^1,u^2)}{\partial u^a} \Bigg|_{u^a =U^a(s)}.$}  It is this constraint that identifies the frame as the Darboux frame, and in turn identifies the curvatures associated with
the frame as $\kappa_g$, $\kappa_n$ and $\tau_g$. This may appear to represent an overkill: but--as we will see--if this constraint is overlooked, one runs into mathematical inconsistencies which indicate that something is amiss.
\vskip1pc \noindent
We thus construct the modified functional $H_C(s, U^a,\kappa_g, \kappa_n, \tau_g, \mathbf{Y},{\bf T}, \mathbf{L}, \mathbf{N},\mathbf{F}, \bm{\lambda}, \bm{\Lambda}, \lambda_{IJ}, {\cal H}_g, {\cal H}_n,{\cal
T}_g)$:
\begin{eqnarray} \label{Hc}
H_C & = & H(\kappa_g,\kappa_n,\tau_g,V) + \int ds\, \mathbf{F}
\cdot (\mathbf{T} - \mathbf{Y}') \nonumber\\
&&  + \int ds  \,\, \bm{\lambda}(s) \cdot \left[ \mathbf{Y}(s) - \mathbf{X} (U^a(s)) \right] + \int
ds\, \mathbf {\Lambda}\cdot (\mathbf{N}- \mathbf{n}(U^a(s)))\nonumber\\
&& + \frac{1}{2} \int ds\,\lambda_\mathbf{TT} (\mathbf{T} \cdot
\mathbf{T}-1)  + \frac{1}{2} \int ds \lambda_\mathbf{LL}
(\mathbf{L} \cdot \mathbf{L}-1)  + \frac{1}{2} \int ds\, \lambda_\mathbf{NN}
(\mathbf{N} \cdot \mathbf{N}-1) \nonumber\\
&&+ \int ds \,\lambda_\mathbf{TL} \mathbf{T} \cdot \mathbf{L}  +
\int ds \, \lambda_\mathbf{LN} \mathbf{L}
\cdot \mathbf{N}  +\int ds \, \lambda_\mathbf{NT}\mathbf{N} \cdot \mathbf{T}  \nonumber\\
&&+ \int ds\, {\cal H}_g  (\mathbf{T}' \cdot \mathbf{L}-\kappa_g)
+ \int ds\, {\cal T}_g (\mathbf{L}' \cdot \mathbf{N} - \tau_g)
+\int ds\,  {\cal H}_n (\mathbf{N}' \cdot \mathbf{T}-\kappa_n) \,.
\end{eqnarray}
The introduction of appropriate Lagrange multipliers frees all three Darboux frame vectors and the curvatures defined by their rotation to be varied independently.\footnote{This may appear,  at  first sight, to   be an overly roundabout approach to the problem. We will see, however, that there is a direct payoff: the shape equation gets cast directly  in terms of the evolution of the Euclidean vector $\mathbf{F}$ along the curve.  If there is a conservation law lurking, it will be sniffed out;  boundary conditions become transparent. The first point was clear already in reference \cite{Auxiliary},  the latter in reference \cite{Contact}, both in the context of membranes.} Note that the Lagrange multiplier $\lambda_\mathbf{TT}$ ensures that the parameter $s$ is arc-length.
\vskip1pc \noindent
The variable $\mathbf{Y}$ appears only in the second and third terms in Eq. (\ref{Hc}). The variation of $H_C$ with respect to
$\mathbf{Y}$ is thus given by
\begin{equation}
\delta_\mathbf{Y} H_C = \int ds \left(\mathbf{F}' + {\bm \lambda}\right) \cdot \delta \mathbf{Y}\,.
\end{equation}
Thus, in equilibrium, one finds that
\begin{equation} \label{Fprime}
\mathbf{F}' = - \bm{\lambda}\,.
\end{equation}
The fate of boundary terms arising from total derivatives will be addressed in Sect. \ref{BCs}.  In addition, in Sect. \ref{sect:translations} it will be seen that $\mathbf{ F}$ is the tension along the curve. It is not conserved however: external forces  associated with the contact constraint acting along the curve are captured by the multiplier ${\bm \lambda}$ \cite{LandauLifshitz}. Surface curvature breaks the translational invariance of $H$.
\vskip1pc \noindent
The variables $U^a$ appear in the third and fourth terms. The corresponding variation of $H_C$ with respect to $U^a(s)$ is given by
\begin{equation}
\delta_U H_C = - \int ds\,\left( \bm{\lambda} \cdot \mathbf{e}_a (U^a) + K_{ab} \Lambda^b -  \partial_a V \right)\, \delta U^a\,.
\end{equation}
Here we have defined $\Lambda^a= \mathbf{\Lambda}\cdot {\bf e}^a$, and made use of the Weingarten surface structure equations, capturing the definition of the extrinsic curvature, $\partial_a \mathbf{n}=K_{a}^{\phantom{a}b}\, \mathbf{e}_b$. Thus, in equilibrium,
\begin{equation} \label{lamLam}
\bm{\lambda}\cdot \mathbf{e}_a = - K_{ab} \Lambda^b + \partial_a V\,;
\end{equation}
the force on the equilibrium curve generally does not act orthogonally to the surface: $\bm{\lambda} \cdot \mathbf{e}_a\ne 0$. Notice that the potential depends on the trajectory through its dependence on the functions $U^a$.
\vskip1pc \noindent
Since $H$ is a scalar under reparametrization, the tangential projection of $ \mathbf{F}'$ must vanish,
\begin{equation} \label{FprT0}
 \mathbf{F}' \cdot \mathbf{T} = 0\,,
\end{equation}
whether the curve is in equilibrium or not \cite{Hamforcurves}.\footnote{That it is not identically satisfied here is because we have broken the manifest reparametrization invariance of the problem by the explicit introduction of parametrization by arc-length.} A consequence is that the projection along $\mathbf{T}$ of the external force must vanish: $\mathbf{T}\cdot \bm{\lambda}=0$, so that from Eq. (\ref{lamLam}) one gets $t^a K_{ab} \Lambda^b = V'$, which can be alternatively cast in the index-free form
\begin{equation} \label{LamTLamL}
\kappa_n \Lambda^{\mathbf{T}}  -\tau_g \Lambda^{\mathbf{L}} = V'\,,
\end{equation}
where $\Lambda^{\mathbf{T}}= t_a \Lambda^a$ and $\Lambda^{\mathbf{L}}= l_a \Lambda^a$ are the projections of the
surface vector $\Lambda^a$ along $\mathbf{T}$ and $\mathbf{L}$ respectively. One can thus express the tangential component of $\bm{\lambda}$ along $\mathbf{L}$ in the form ($\nabla_\mathbf{L}  V = l^a \partial_a V$)
\begin{equation} \label{eq:lampar}
\lambda^\mathbf{L} = \bm{\lambda} \cdot \mathbf{L}= \frac{1}{\kappa_n} (-{\cal K}_G \, \Lambda^\mathbf{L} + \tau_g V')
  + \nabla_\mathbf{L} V \,.
\end{equation}
Here we have used the identities for the normal curvature and geodesic torsion in terms of the relevant projections of the extrinsic curvature tensor, given by Eq. (\ref{kaptaudef}); we have also used the completeness of the tangent vectors $\mathbf{T}$ and $\mathbf{L}$ on the surface: $t^a t^b + l^al^b = g^{ab}$, to express the normal curvature along the orthogonal direction in the form: $l^a l^b K_{ab}= K - \kappa_n$; the definition of the Gaussian curvature as a two-dimensional determinant then identifies
\begin{equation} \label{KGdef}
 {\cal K}_G= 1/2 (K^2 - K^{ab} K_{ab}) = \kappa_n  (K - \kappa_n) -\tau_g^2\,,
 \end{equation}
 of which, Eq. (\ref{KGkntg}) is a special case.
\vskip1pc \noindent
The two Lagrange multipliers $\Lambda^{\mathbf{L}}$ and $\lambda^\mathbf{N}$ remain undetermined. To anticipate, $\Lambda_{\mathbf{L}}$ will appear in the homogeneous EL equation for $\mathbf{N}$; this equation will then fix this multiplier, and with it the tangential force--through Eq. (\ref{eq:lampar})--completely in terms of the local geometry.
\vskip1pc \noindent
From the variation of $H_C$ with respect to $\kappa_g$, $\kappa_n$ and $\tau_g$ we readily determine the Lagrange
multipliers  expressing their definitions in terms of the Darboux basis vectors by Eqs. (\ref{Lambdatnnt}).
\vskip1pc \noindent
Now $H_C$ is stationary under variations with respect to $\mathbf{T}$, $\mathbf{N}$ and $\mathbf{L}$
respectively when
\begin{subequations}
\begin{eqnarray}
\mathbf{F} + \lambda_\mathbf{TT} \mathbf{T} + \lambda_\mathbf{TL} \mathbf{L} + \lambda_\mathbf{NT} \mathbf{N} +{\cal H}_n  \mathbf{N}' - ({\cal H}_g \mathbf{L})' &=& 0\,, \label{TEL0}\\
\lambda_\mathbf{TL} \mathbf{T} + \lambda_\mathbf{LL} \mathbf{L} + \lambda_\mathbf{LN} \mathbf{N}
+ {\cal H}_g \mathbf{T}' - ({\cal T}_g \mathbf{N})' & = & 0\,, \label{LEL0}\\
\mathbf{\Lambda} + \lambda_\mathbf{NT} \mathbf{T}  + \lambda_\mathbf{LN} \mathbf{L} + \lambda_\mathbf{NN} \mathbf{N} + {\cal T}_g  \mathbf{L}' - ({\cal H}_n  \mathbf{T})' & = & 0\,. \label{NEL0}
\end{eqnarray}
\end{subequations}
Substituting the structure equations (\ref{Darbouxequations}) into Eqs. (\ref{TEL0})-(\ref{LEL0}), one obtains
\begin{subequations}
\begin{eqnarray}
\mathbf{F}+(\lambda_\mathbf{TT} + \kappa_g {\cal H}_g + \kappa_n {\cal H}_n ) \mathbf{T}
+ (\lambda_\mathbf{TL} -  {\cal H}_g' - \tau_g  {\cal H}_n ) \mathbf{L}
+ \left(\lambda_\mathbf{NT} - \tau_g {\cal H}_g \right) \mathbf{N}&=&0\,, \label{TEL}\\
(\lambda_\mathbf{TL} - \kappa_n {\cal T}_g )\mathbf{T}
+ (\lambda_\mathbf{LL} + \kappa_g {\cal H}_g  + \tau_g {\cal T}_g ) \mathbf{L}
+(\lambda_\mathbf{LN}  - {\cal T}_g' - \kappa_n {\cal H}_g ) \mathbf{N} & = & 0\,, \label{LEL}\\
\mathbf{\Lambda} +(\lambda_\mathbf{NT} - {\cal H}'_n - \kappa_g {\cal T}_g )\mathbf{T}
 + \left( \lambda_\mathbf{LN} - \kappa_g {\cal H}_n \right) \mathbf{L}
+ (\lambda_\mathbf{NN}+\kappa_n {\cal H}_n + \tau_g {\cal T}_g ) \mathbf{N}& = & 0\,. \label{NEL}
\end{eqnarray}
\end{subequations}
The first equation expresses the tension $\mathbf{F}$ at each point along the curve as a linear combination of the Darboux
vectors; the linear independence of these vectors implies that each of the six coefficients appearing in Eqs. (\ref{LEL}) and (\ref{NEL}) must vanish, thus:
\begin{subequations} \label{ecsnfs}
\begin{eqnarray}
\lambda_\mathbf{TL} - \kappa_n {\cal T}_g & = &0, \label{ecsbfsb1}\\
\lambda_\mathbf{LL} + \kappa_g {\cal H}_g  + \tau_g {\cal T}_g & = & 0\,, \label{ecsbfsb2}\\
\lambda_\mathbf{LN} - {\cal T}'_g - \kappa_n {\cal H}_g & = & 0,  \label{ecsbfsb3}\\
\Lambda^\mathbf{T} + \lambda_\mathbf{NT} - {\cal H}'_n - \kappa_g {\cal T}_g &= &0\,, \label{ecsnfsb1}\\
\Lambda^\mathbf{L} + \lambda_\mathbf{LN} - \kappa_g {\cal H}_n & = & 0\,, \label{ecsnfsb2}\\
\Lambda^\mathbf{N} + \lambda_\mathbf{NN} + \kappa_n {\cal H}_n + \tau_g {\cal T}_g & = & 0\,. \label{ecsnfsb3}
\end{eqnarray}
\end{subequations}
Eqs. (\ref{ecsbfsb1}), (\ref{ecsbfsb2}) and (\ref{ecsbfsb3}) determine three of the multipliers directly in terms of $\kappa_g$, $\kappa_n$ and $\tau_g$:
\begin{subequations} \label{lambda}
\begin{eqnarray}
\lambda_\mathbf{TL} &= & \kappa_n {\cal T}_g\,,\label{lambdalt}\\
\lambda_\mathbf{LL} & = & - \kappa_g {\cal H}_g -  \tau_g {\cal T}_g\,, \label{lambdal}\\
\lambda_\mathbf{LN} & = & {\cal T}'_g + \kappa_n {\cal H}_g \,. \label{lambdanl}
\end{eqnarray}
\end{subequations}
Eq. (\ref{lambdanl}) together with Eq. (\ref{ecsnfsb2}) completely determines $\Lambda^\mathbf{L}$
\begin{equation} \label{LamLng}
\Lambda^\mathbf{L} = - {\cal T}_g' + \kappa_g {\cal H}_n - \kappa_n {\cal H}_g \,.
\end{equation}
This vanishes for the Frenet energy, given by Eq. (\ref{HBF}), and only for this energy.
\vskip1pc \noindent
In turn Eq. (\ref{LamLng}), determines $\Lambda^\mathbf{T}$ through Eq. (\ref{LamTLamL})
\begin{equation} \label{eq:lamT}
\Lambda^\mathbf{T} = \frac{1}{\kappa_n} \left(\tau_g (- {\cal T}_g' +\kappa_g {\cal H}_n- \kappa_n {\cal H}_g)+V' \right)\,.
\end{equation}
\vskip1pc \noindent
Together Eqs. (\ref{eq:lampar}) and (\ref{LamLng}), also completely determine the tangential force
$\lambda^\mathbf{L}$ in terms of the geometry:
\begin{equation} \label{eq:lamL}
\lambda^\mathbf{L} = \frac{ K_G}{\kappa_n} \, ({\cal T}_g' + \kappa_n {\cal H}_g - \kappa_g {\cal H}_n ) + \frac{\tau_g V' }{\kappa_n} + \nabla_\mathbf{L} V \,.
\end{equation}
Using Eq. (\ref{eq:lamT}) for $\Lambda^\mathbf{T}$ in Eq. (\ref{ecsnfsb1}) gives
\begin{equation}
\lambda_\mathbf{NT} = {\cal H}'_n + \tau_g\, {\cal H}_g  + \kappa_g {\cal T}_g  + \frac{1}{\kappa_n} \left( \tau_g ({\cal T}_g' -\kappa_g {\cal H}_n)  - V' \right) \,.
\end{equation}
We  have now determined all of the relevant Lagrange multipliers,\footnote{Using Eq. (\ref{ecsnfsb3}) one determines $\lambda_\mathbf{NN} = -\Lambda^\mathbf{N} - \kappa_n {\cal H}_n - \tau_g {\cal T}_g$.  One can set $\Lambda^\mathbf{N}=0$ without any loss of generality.  While $\lambda_\mathbf{NN}$ itself does not feature in the physical description of the curve, had the corresponding constraint been overlooked one would have run into an inconsistency.}  save one:
$\lambda_\mathbf{TT}$. The tension along the curve $\mathbf{F}$ given by Eq. (\ref{TEL0}) is now determined modulo this multiplier.  To complete the description, one  appeals again to the statement of reparametrization invariance of $H$, Eq.        (\ref{FprT0}), which along with Eq. (\ref{TEL}) implies that
\begin{equation}  \label{fpdpt}
(\lambda_\mathbf{TT} + \kappa_g {\cal H}_g + \kappa_n {\cal H}_n )' + \kappa_g {\cal H}_g' + \kappa_n {\cal H}_n ' + \tau_g {\cal T}_g' - V' =0 \,.
\end{equation}
Eq. (\ref{fpdpt}) can be integrated to express $\lambda_\mathbf{TT}$ as a linear functional of ${\cal H}$:
\begin{equation}
 \lambda_\mathbf{TT} = {\cal H} -2 (\kappa_g {\cal H}_g + \kappa_n {\cal H}_n )
 - \tau_g {\cal T}_g + V +  c \,,
\end{equation}
where $c$ is a constant. Thus, modulo this constant, the vector $\mathbf{F}$--conserved or not--is now completely determined by the curve itself:
\begin{equation} \label{forceDB}
\mathbf{F} = F^\mathbf{T} \, \mathbf{T} + F^\mathbf{L}\,  \mathbf{L} + F^\mathbf{N} \, \mathbf{N}\,,
\end{equation}
where $F^\mathbf{T}, F^\mathbf{L}$ and $F^\mathbf{N}$ are given by Eqs. (\ref{FTNL}). Using Eq. (\ref{Fprime}),  $\mathbf{F}$ satisfies
\begin{equation} \label{eq:Fprime}
\mathbf{F}' =  - \lambda^\mathbf{L}\,\mathbf{L} - \lambda^\mathbf{N} \,\mathbf{N}\,.
\end{equation}
where $\lambda^\mathbf{L}$ is given by Eq. (\ref{eq:lamL}), and $\lambda^\mathbf{N} = \bm{\lambda} \cdot \mathbf{N}$. There is no projection onto $\mathbf{T}$ on account of Eq. (\ref{FprT0}). The EL equation  is then given by $\epsilon_\mathbf{L} = \mathbf{F}^{'} \cdot \mathbf{L} + \lambda^\mathbf{L}=0$, where $\mathbf{F}' \cdot \mathbf{L}= F^\mathbf{L}{}' +\kappa_g F^\mathbf{T}-\tau_g F^\mathbf{N}$, which reproduces Eq. (\ref{EL}). The magnitude of the normal force is given by $\lambda^\mathbf{N} = - \mathbf{F}' \cdot \mathbf{N} =  - F^\mathbf{N}{}' + \kappa_n \, F^\mathbf{T} - \tau_g \, F^\mathbf{L}$,  which reproduces Eq. (\ref{NF}).
\vskip1pc \noindent
Before examining the general structure,  it is useful  to first confirm that this framework reproduces known results for bound Euler elastic curves.

\subsection{Bound Euler-Elastic curves}

The bending energy defined by Eq. (\ref{HBF}), can be decomposed as
\begin{equation} \label{HBkgkn}
 H_B = 1/2 \int ds (\kappa^2_g + \kappa^2_n)\,,
\end{equation}
thus, ${\cal H}_g=\kappa_g$, ${\cal H}_n= \kappa_n$, with ${\cal T}_g=0$ and $V=0$. In addition, there are no tangential forces: $\lambda^\mathbf{L}$ given by Eq. (\ref{eq:lamL0}) vanishes due to the  quadratic dependence on curvature as well as the symmetry between
$\kappa_g$ and $\kappa_n$. Note that the symmetry, itself,  is not enough.\footnote{More generally, if the dependence of ${\cal H}$ on $\kappa_g$ and $\kappa_n$ can be collected into a dependence on $\kappa$, $\lambda^\mathbf{L}$ will vanish.}
The EL equation  (\ref{EL}) reduces to
\begin{equation} \label{ELEulerelasta}
\varepsilon_\mathbf{L} = \kappa_g'' + \frac{\left(\kappa^2_n \tau_g\right)'}{\kappa_n} + \kappa_g \left(\frac{\kappa_g^2 + \kappa_n^2}{2} - \tau_g^2-c\right) = 0\,.
\end{equation}
This agrees with the equation first derived in reference \cite{MannNick} and, using an approach adapted to the Frenet basis, in \cite{GuvVaz}. In general,  geodesics with $\kappa_g =0$ are not solutions.
\vskip1pc \noindent
For this energy, the magnitude of the normal force $\lambda_\mathbf{N}$, given by (\ref{NF}), transmitted to the surface assumes the form
\begin{equation}
\label{NFEuler}
\lambda^\mathbf{N} = \kappa_n'' - \frac{\left(\kappa_g^2 \tau_g \right)'}{\kappa_g} + \kappa_n \left( \frac{\kappa_g^2 + \kappa_n^2}{2} - \tau_g^2 -c\right) \,,
\end{equation}
which again reproduces the result first obtained in \cite{GuvVaz}. The normal force depends non-trivially on the location on the surface.

\subsection{Boundary conditions on an open curve}\label{BCs}

Four terms were consigned to the end points of the curve in the variational principle. Modulo the EL equations, the variation of the Hamiltonian (\ref{Hc}) is given by $\delta H_C = - \int ds\,\delta Q'$, where the boundary term $\delta Q$ is identified as
\begin{equation} \label{ecQ}
\delta Q = \mathbf{F} \cdot \delta \mathbf{Y} - {\cal H}_g  \mathbf{L}  \cdot \delta \mathbf{T} - {\cal T}_g \mathbf{N} \cdot \delta \mathbf{L} - {\cal H}_n \mathbf{T} \cdot \delta \mathbf{N}  \,.
\end{equation}
Several different boundary possibilities are physically relevant.

\subsubsection{Fixed ends}

If the end points are fixed then the first and last terms in (\ref{ecQ}) vanish: $\delta \mathbf{Y}=0$ and $\delta \mathbf{N}=0$.  If the tangent is also fixed, then both middle terms vanish as well, for $\delta \mathbf{T}=0$ and consequently $\delta \mathbf{L}=0$.\footnote {Note that $\delta \mathbf{Y}=0$ implies that both $\Psi_\|$ and $\Psi_\perp$ defined by Eq. (\ref{delalphaeulelonsurf}) vanish. Then $\delta \mathbf{T} = \Psi'_\perp \, \mathbf{L}$. So fixing the tangent at the end point is equivalent to setting $\Psi_\perp'=0$ there.} If $\mathbf{T}$ is not fixed one requires ${\cal H}_g=0$.
\vskip1pc \noindent
The boundary term also vanishes for a periodic curve such as a spiral on a helicoid.

\subsubsection{One or two free ends}

The constraint that the curve $\mathbf{Y}$ lie on the surface implies that, at the boundary, its variation $\delta \mathbf{Y}$ must be tangential:  we decompose it with respect to the tangent basis $\{\mathbf{T},\mathbf{L}\}$,\footnote{Note that whereas a tangential deformation at interior points can always be identified with a  reparametrization, at a boundary it cannot. On the boundary, it changes the length of the curve.}\footnote{ Equivalently, $\delta U^a = \Psi_\| t^a + \Psi_\perp l^a$.}
\begin{equation} \label{delalphaeulelonsurf}
\delta \mathbf{Y}=\Psi_\|  \mathbf{T} + \Psi_\perp \mathbf{L}\,.
\end{equation}
The requirement that $s$ remain arc-length under variation implies that variation and differentiation with respect to arc length commute: $\delta \mathbf{T} = (\delta \mathbf{Y})' $, or equivalently $\mathbf{T}\cdot\delta \mathbf{T}=0$. This implies the constraint
\begin{equation} \label{isomcond}
 \Psi'_\parallel - \kappa_g \Psi_\perp = 0
\end{equation}
on the two component of the variation field. Thus
\begin{subequations}
\begin{eqnarray}
\delta \mathbf{T} &=& (\Psi'_\perp + \kappa_g \Psi_\|) \mathbf{L} + (\tau_g \Psi_\perp - \kappa_n \Psi_\|)\mathbf{N}\,, \\
\delta \mathbf{N} &=& - (\tau_g \Psi_\perp - \kappa_n \Psi_\|)\mathbf{T} + ((K - \kappa_n) \Psi_\perp - \tau_g \Psi_\|) \mathbf{L}\,,
\end{eqnarray}
\end{subequations}
and $\delta \mathbf{L}$ follows from the orthonormality of the Darboux frame. Notice that $\delta\mathbf{N}$ does not involve derivatives of the deformation scalars, consistent with its vanishing when $\mathbf{Y}$ is fixed.\footnote{
The corresponding changes in the Darboux curvatures follow
\begin{subequations}
\begin{eqnarray}
\delta \kappa_g &=& \Psi''_\perp + {\cal K}_G \Psi_\perp  + (\kappa_g \Psi_\parallel)' \,,\\
\delta \kappa_n &=& - 2 \tau_g \Psi'_\perp - (\tau'_g + \kappa_g (K - \kappa_n)) \Psi_\perp  + (\kappa_n \Psi_\parallel)' \,,\\
\delta \tau_g &=& - (K \Psi_\perp)' + 2 \kappa_n \Psi'_\perp + (\kappa'_n + \kappa_g \tau_g) \Psi_\perp  + (\tau_g \, \Psi_\parallel)' \,.
\end{eqnarray}
\end{subequations}
The constraint (\ref{isomcond}) permits the $\Psi_\|$ dependence to be absorbed into a total derivative. It is simple to confirm that these expressions provide an alternative derivation of the single EL equation (\ref{EL}), as well as the boundary conditions (\ref{eulelsupbouncond}). What is missing is the underlying structure of the EL equation in terms of a tension vector, provided by the constrained variational approach provided earlier as well as the identification of the normal forces.}
The boundary term  (\ref{ecQ}) now reads
\begin{equation} \label{deltaQfree}
\delta Q = - {\cal H}_g\, \Psi'_\perp + \left( {\cal H}_g' + 2 \tau_g  {\cal H}_n + (K - 2 \kappa_n) {\cal T}_g \right) \, \Psi_\perp  -( {\cal H} + V + c) \, \Psi_\| \,.
\end{equation}
Vanishing $\delta Q$ at the end point for arbitrary $\Psi'_\perp$, $\Psi_\perp$ and $\Psi_\|$  implies the boundary conditions
\begin{equation} \label{eulelsupbouncond}
{\cal H}_g = 0 \,, \quad {\cal H}'_g = - 2 \tau_g {\cal H}_n - (K - 2 \kappa_n) {\cal T}_g  \,,\quad {\cal H} = - V - c \,,
\end{equation}
at free end-points. For a bound Euler elastica, it implies that the curve must be geodesic at a free end, so these boundary conditions read
 \begin{equation} \label{eulerbc}
\kappa_g=0\,, \quad \kappa_g'  = - 2 \tau_g \kappa_n  \,,\quad {\cal H} +  c =0 \,.
\end{equation}
The last condition, in turn, implies $\kappa_n^2 =-2c$ so that $c<0$. On a sphere, the only equilibrium states consistent with Eq. (\ref{eulerbc}) are geodesic arcs.

\section{Examples of surface biased bending energies}

We now examine various energies of special interest.

\subsection{Energies dependent only on surface intrinsic geometry}

Consider an elastic curve on a surface with the intrinsically defined energy
\begin{equation} \label{kgKG}
H = \frac{1}{2} \int ds \, \left(\kappa_g^2 - \gamma {\cal K}_G\right)\,,
\end{equation}
involving a potential proportional to the local surface Gaussian curvature, ${\cal K}_G$; $\gamma$ is a constant.\footnote
{
An interesting limit in its own right is one in which $H$ depends only on the potential, so that $H = \int ds \, {\cal K}_G$.  One may be interested in identifying curves that avoid regions of large Gaussian curvature, so that one minimizes (or maximizes)  the total Gaussian curvature along its length.}  Higher dimensional relativistic brane world  analogues of (\ref{kgKG}), with energy replaced by the action, have been considered in some detail by Armas \cite{Armas}.
\vskip1pc \noindent
In this case one has ${\cal H}_g = \kappa_g$, ${\cal H}_n ={\cal T}_g = 0$ and $V = - \gamma {\cal K}_G /2 $;
thus Eq. (\ref{EL}) can be cast completely in terms of the intrinsic geometry, independent of the extrinsic curvature\footnote{Recall that both the geodesic curvature and the Gaussian curvature are invariant under isometry.}
\begin{equation} \label{kgKGEL}
\kappa_g''   + \kappa_g \left( \frac{\kappa^2_g}{2} + \left(1 + \frac{\gamma}{2}\right) {\cal K}_G - c \right) = \frac{\gamma}{2} \nabla_\mathbf{L} {\cal K}_G\,.
\end{equation}
If, in addition ${\cal K}_G$ is constant, geodesics occur as solutions.
\vskip1pc \noindent
Despite the intrinsic nature of the Euler Lagrange equation,  (\ref{kgKGEL}), the magnitude of the normal force binding the curve to the surface, given by Eq. (\ref{NF}), is generally non-vanishing and does depend explicitly on the extrinsic curvature:
\begin{equation} \label{NFK}
\lambda^\mathbf{N} = \frac{\gamma}{2} \left(\frac{{\cal K}'_G}{\kappa_n}\right)' - \tau_g \kappa_g' + \kappa_n \left(\frac{\kappa^2_g}{2} + \frac{\gamma}{2} {\cal K}_G - c \right) \,.
\end{equation}
This is true even when ${\cal K}_G$ is constant and along geodesics. In this case,  the force is proportional to $\kappa_n$.
\vskip1pc \noindent
If the surface is a round sphere of radius $R_0$,  with ${\cal K}_G=1/R_0^2$, $\kappa_n=1/R_0$ and $\tau_g=0$, the two energies   (\ref{HBkgkn}) and (\ref{kgKG}) coincide if $\gamma=-1$. Consequently Eq. (\ref{kgKGEL}) coincides with Eq. (\ref{ELEulerelasta}), and Eq. (\ref{NFK}) with Eq.(\ref{NFEuler}).  The force depends only on the local value of the geodesic curvature.
\vskip1pc \noindent
On a round sphere, the extrinsic geometry like the metric is homogeneous and isotropic. There are, however,  geometries locally isometric to a sphere, which are non-trivially embedded. Examples are provided by the axially symmetric surfaces of constant positive Gaussian curvature obtained by removing or adding a wedge spanned by two meridians from a sphere ;  both geometries display conical singularities
\cite{DoCarmo, Gray}.  Whereas geodesics on a round sphere are sections of closed great circles, they generally do not close in these isometric geometries; and while $\lambda^\mathbf{N} $ is constant along the great circles on a round sphere, it will not be along geodesics in one of these geometries.

\subsubsection{Energy symmetric in curvatures and torsion}

Let us now examine the energy symmetric in the curvatures and the torsion,
\begin{equation}  \label{kkt}
H = \frac{1}{2} \int ds\, \left( \kappa_g^2 + \kappa_n^2 + \tau_g^2 \right)\,.
\end{equation}
On first inspection, this energy appears very different from the intrinsically defined Eq. (\ref{kgKG}) and, in general, it is. However,  using the identity Eq. (\ref{KGdef}), it is evident that the two-energies (\ref{kkt}) and (\ref{kgKG}) coincide on minimal surfaces when $\gamma=1$.  Using the identities ${\cal H}_g=\kappa_g$, ${\cal H}_n= \kappa_n$, ${\cal T}_g=\tau_g$ and $V=0$, the EL equation  (\ref{EL})  reduces to
\begin{equation} \label{kktEL}
\kappa_g'' + \kappa'_n \tau_g + \left(K -\kappa_n\right) \tau'_g + \kappa_g \left( {\cal H}  -  c \right) = 0 \,.
\end{equation}
where we have used Eq. (\ref{KGdef}) to replace ${\cal K}_G$ in favor or the Darboux curvatures and $K$. The magnitude of the normal force (\ref{NF}) is given by
\begin{equation} \label{fpdpn}
\lambda^\mathbf{N} = \kappa''_n +\left(\frac{\tau_g  \tau_g'}{\kappa_n}\right)' - \tau_g \kappa _g' + \kappa_n \left({\cal H} - c \right) \,.
\end{equation}
On a minimal surface, the EL equation (\ref{kktEL}), unlike its counterpart (\ref{kgKGEL}), appears to depend explicitly on the extrinsically defined $\kappa_n$ and $\tau_g$.  As we will show, however, on such a surface Eq. (\ref{kktEL}) can be cast in a form that is manifestly independent of the extrinsic geometry.   First note that Eq. (\ref{kktEL}) can be cast, using Eq. (\ref{KGdef}),  as
\begin{equation} \label{kktELK0}
\kappa_g^{''}  + \kappa_g \left( \frac{1}{2} \kappa^2_g - \frac{1}{2} {\cal K}_G - c \right) =  \kappa_n \tau_g' - \kappa_n' \tau_g\,,
\end{equation}
whereas Eq. (\ref{kgKGEL}) with $\gamma=1$ can be written as
\begin{equation}
\kappa_g^{''}  + \kappa_g \left( \frac{1}{2} \kappa^2_g - \frac{1}{2} {\cal K}_G - c \right) =  \frac{1}{2} \nabla_\mathbf{L} {\cal K}_G - 2 \kappa_g {\cal K}_G\,.
\end{equation}
On a minimal surface,  however, it is shown in \ref{AppCM} that
\begin{equation} \label{kKkkt}
\nabla_\mathbf{L} K_G/2 = \kappa_n \tau_g' - \tau_g \kappa_n' + 2 \kappa_g {\cal K}_G\,.
\end{equation}
Using this identity,
Eqs. (\ref{kktEL}) and (\ref{kgKGEL}) with $\gamma=1$ indeed coincide on such  surfaces.
\vskip1pc \noindent
As further illustration of the flexibility of this framework, we examine two futher examples, special cases  of which  have been treated in the recent literature.

\subsubsection{Energy quadratic in curvature and linear in geodesic torsion} \label{crandisken}

The energy ${\cal H} = \kappa^2/2 + \nu \tau_g$ was proposed in Ref. \cite{Kaplan} in order to model the energy of boundaries of axially symmetric cranellated disks, where the geodesic torsion term accounts for their microscopic chirality. For this energy one has that ${\cal T}_g = \nu$, so the EL derivative in (\ref{ELEulerelasta}) has the additional contribution $-\nu  (\kappa'_n -  \kappa_g \tau_g)$.
Likewise, the magnitude of normal force given by (\ref{NFEuler}) has the additional term $\nu (\kappa_g' +  \kappa_n \tau_g )$.

\subsubsection{Barros Garay energy}

Recently Barros and Garay considered the stationary states of an energy depending only on the normal curvature ${\cal H} = \kappa^N_n/N$, and in particular, for $N = 2$,  for curves on space forms \cite{BarrosGaray}. It is simple to show that Eqs. (\ref{EL}) and (\ref{NF}) reduce to
\begin{equation} \label{ELBar}
\varepsilon_\mathbf{L} =  2 \, \tau_g \, (\kappa^{N-1}_n)' + \kappa^{N-1}_n \, \tau'_g + \kappa_g \left( \left(2 -\frac{1}{N}\right) \kappa^N_n - K \, \kappa^{N-1}_n - c \right)=0\,,
\end{equation}
and
\begin{equation} \label{NFBar}
 \lambda^\mathbf{N} = (\kappa^{N-1}_n)'' - \left( \kappa_g \, \tau_g\, \kappa^{N-2}_n \right)'  + \kappa_n \left (\left(1-\frac{1}{N}\right) \,  \kappa^N_n -\tau^2_g \kappa^{N-2}_n  -  c \right) \,.
\end{equation}
Except for a minus sign arising from the different convention used for $\kappa_n$ and the constant $c$ fixing total length, the EL eq. (\ref{ELBar}) coincides with expression ($2.19$) presented in proposition $1$ of Ref. \cite{BarrosGaray}.

\section{Residual Euclidean invariance}\label{Euclid}

\subsection{Translations} \label{sect:translations}

In general,  the change in the energy along any section of a bound curve under a deformation is given by
\begin{equation}
\delta H_C =\int ds \, (\mathbf{F}' + \bm{\lambda}) \cdot \delta\mathbf{Y}  - \int ds \, \delta Q'
\,,\end{equation}
where $\delta Q$ is given by Eq. (\ref{ecQ}).
In equilibrium,
$\mathbf{F}' = - \bm{\lambda}$.
\vskip1pc \noindent
Under an infinitesimal translation of the surface, $\delta \mathbf{Y} = \delta \mathbf{c}$, the Darboux vectors along the bound curve do not change, so that $\delta \mathbf{T} = 0$, $\delta \mathbf{L} = 0$, and  $\delta \mathbf{N} = 0$.
The boundary term  is then given by $\delta Q =  \delta \mathbf{c} \cdot \mathbf{F}$. In equilibrium,
\begin{equation}
\delta H_C = -\delta \mathbf{c} \cdot \int ds \,\mathbf{F}' =  \delta \mathbf{c} \cdot \int ds \,\bm \lambda\,.
\end{equation}
Thus  the vector ${\bf F}$ is identified as the tension along the curve and $\int ds {\bm \lambda}$ represents the total force on the curve.
\vskip1pc \noindent
Let the surface geometry be a generalized cylinder (not necessarily circular) so that $H$ is invariant under translation along the $\mathbf{k}$ direction (say).
Then the component
\begin{equation}
F^Z = \mathbf{F} \cdot \hat{\bf k} \,,
\end{equation}
is constant.

\subsection{Rotations} \label{sect:rotations}

Under an infinitesimal rotation of the surface, $\delta \mathbf{Y} = \delta \bm{\omega} \times \mathbf{Y}$ the Darboux vectors change accordingly:
$\delta \mathbf{T} = \delta \bm{\omega} \times \mathbf{T}$, and similarly for $\mathbf{L}$ and $\mathbf{N}$.
The corresponding boundary term (\ref{ecQ}) is given by
\begin{equation}
\delta Q =  \delta \bm{\omega} \cdot \mathbf{M}\,,
\end{equation}
where the torque of the curve about the chosen origin is given by
\begin{equation} \label{eq:MFS}
\mathbf{M}= \mathbf{Y} \times \mathbf{F} + \mathbf{S}\,,
\end{equation}
with
\begin{equation} \label{S}
\mathbf{S} = - {\cal T}_g \mathbf{T} - {\cal H}_n \mathbf{L} - {\cal H}_g \mathbf{N}\,.
\end{equation}
The first term appearing in $\mathbf{M}$ is the torque with respect to the origin due to the force $\mathbf{F}$  acting on a
segment of the curve; $\mathbf{S}$ is the bending moment originating in the curvature dependence of  the bending energy,
and is translationally invariant. If ${\cal H}= \kappa^2/2$, one reproduces the expression
\begin{equation} \label{SDarboux}
\mathbf{S} =  - \kappa_n \, \mathbf{L} - \kappa_g \, \mathbf{N} = -\kappa \mathbf{\mathcal B}\,,
\end{equation}
where $\kappa_g = \kappa \cos \theta$ and $\kappa_n = \kappa \sin \theta$ have been used.\footnote{$\theta$ is the angle between the Frenet normal and the Darboux conormal, and $\mathbf{\mathcal B}$ is the Frenet binormal.} On a sphere,  centered on the origin, as described in \cite{GuvVaz}, $\mathbf{M}$ is a constant vector.  However, $\mathbf{M}$ generally will not be  conserved in a bound equilibrium. One has instead
\begin{equation} \label{Mp}
\mathbf{M}' = - \mathbf{Y}\times \, \bm{\lambda} -  \mathbf{N} \times \bm{\Lambda} \,,
\end{equation}
where $\bm \lambda$ is the external force, normal to the curve, defined earlier, Eqs. (\ref{eq:lamL0}) and (\ref{NF}), and $\bm{\Lambda}$ is a vector tangent to the surface, with components defined by  (\ref{LamLng}) and (\ref{eq:lamT}) in \ref{Appderivation}. Thus the source of the torque is given as a sum of two terms: one the moment of the external forces, both normal and tangential to the surface,  associated with the constraint; the second an additional source for the  bending moment.
\vskip1pc \noindent
In the case of bound Euler Elastic curves, Eq. (\ref{Mp})  simplifies. For now $\Lambda^\mathbf{L}=0$ by Eq. (\ref{LamLng}) so that $\bm \Lambda=0$ by Eq. (\ref{LamTLamL}) and  $\lambda^\mathbf{L}=0$ by Eq. (\ref{eq:lampar}).  Thus the source is given by the moments of the normal force $\lambda^\mathbf{N}$ alone. This also lends a physical interpretation for $\bm{\Lambda}$.

\subsubsection{Axially symmetric geometries}

If the surface geometry is  axially symmetric,  $H$ will be  invariant under rotations about the axis of symmetry. As a consequence, the projection of $\mathbf{M}$ along this axis, i.e. $M^Z= \mathbf{M} \cdot \hat{\bf k}$ is conserved. Let $R$ and $Z$ represent the cylindrical polar coordinate and the height along the curve,  adapted to the symmetry, and  $\alpha$ be the angle that the tangent vector to the curve makes with the azimuthal direction (see \ref{axisymm}). One identifies
\begin{equation} \label{MZ}
M^Z =  R \,\left(\cos \alpha \, F^\mathbf{T} - \sin \alpha \,  F^\mathbf{L}\right) + \csc \alpha \, \left(R' \,
{\cal H}_g -  Z' (\cos \alpha  {\cal H}_n + \sin\alpha {\cal T}_g) \right)\,.
\end{equation}
This provides a first integral of the EL equation (\ref{EL}),  one which does not  exactly leap off the page on first inspection. Differentiating it with respect to arc-length one finds,  as shown in  \ref{AppMZp},  that $M^Z{}' = -R \sin  \alpha \,  \varepsilon_\mathbf{L}$ so that in equilibrium $M^Z$ is indeed conserved. Let the surface be parametrized by the length along the meridian and the polar angle $\varphi$. Their values along the curve are determined by
the equations (see  \ref{axisymm})
\begin{subequations}
\begin{eqnarray}
l' &=& \sin\alpha\,, \label{lprime}\\
R\varphi' &=& \cos\alpha\,. \label{phiprime}
\end{eqnarray}
\end{subequations}
The Darboux curvatures and torsion are given by Eqs. (\ref{axikappag}), (\ref{axikappan}) and (\ref{axitaug}) in \ref{axisymm},  or
\begin{equation} \label{axidarboux}
\kappa_g = - \frac{(R\,\cos \alpha)'}{R\sin \alpha}\,; \quad
\kappa_n = \sin^2 \alpha \, C_\perp + \cos^2 \alpha \,C_\parallel\,; \quad
 \tau_g = \sin \alpha \cos \alpha (C_\parallel - C_\perp)\,,
\end{equation}
where $C_\perp$ and $C_\|$ are the meridional and parallel curvatures respectively. Substituting the expressions (\ref{FTNL}) for the components of ${\bf F}$, as well as the identities $Z'= R C_\|   l'= \sin\alpha R C_\parallel$ (use Eq.(\ref{CparCperp}) in \ref{axisymm}) and $2 \cot 2 \alpha \tau_g = 2 \kappa_n - K$ (which follows from the definitions (\ref{axikappan}) and (\ref{axitaug}))
yields the more transparent expression,
\begin{equation} \label{MZaxi}
M^Z = -R  \sin \alpha \left({\cal H}'_g + 2 \tau_g {\cal H}_n + (K - 2 \kappa_n) {\cal T}_g \right) + (R \sin \alpha)' {\cal H}_g - R \cos \alpha ({\cal H} + V+ c)\,.
\end{equation}
For a bound Euler elastic curve, this expression reduces to
\begin{equation} \label{MZaxielastic}
M^Z = -R  \sin \alpha \left(\kappa'_g + 2 \tau_g \kappa_n \right) + R \cos \alpha  \left( \frac{\kappa^2_g - \kappa^2_n}{2}-c \right) + R' \csc \alpha \kappa_g \,.
\end{equation}
The polar radius $R$, as well as the curvatures $C_\|$ and $C_\perp$ appearing in Eq. (\ref{MZaxi}) are given functions of the arc-length along the meridian $l$. In general one can write Eqs. (\ref{lprime}) and Eq. (\ref{MZaxi}) as a pair of coupled ODEs for $\alpha$ and $l$ as functions of $s$: If ${\cal H}$ is quadratic in $\kappa_g$,  Eq. (\ref{MZaxi})  will be linear in second derivatives of $\alpha$, or of the form $\alpha'' = f(\alpha, \alpha', l)$, where $l$ satisfies Eq. (\ref{lprime}).
These two equations provide $R$ and $Z$ as functions of $s$. To complete the construction of the trajectory one uses Eq. (\ref{phiprime}) to determine $\varphi$ as a function of $s$.
\vskip1pc \noindent
Comment on boundary conditions: Suppose that the two ends of a curve of fixed length $L$ are fixed so that the elevations $Z_0$ and $Z_1$, the angle turned  $\Delta \varphi= \varphi_1-\varphi_0$,  as well as the tangent angles at the two points  are given. \footnote {We will suppose that $R$ is a monotonic function of $Z$,  so that $R_0$ and $R_1$ are fixed once $Z_0$ and $Z_1$ are.}  The non-local constraints on the length and the azimuthal angle turned  completely determine the  two free parameters appearing in Eq. (\ref{MZaxi}). Solutions on a catenoid will be studied in detail elsewhere.
\vskip1pc \noindent
If free-boundary conditions are relevant, the solution is very different. Comparing Eqs. (\ref{eulelsupbouncond}) and (\ref{MZaxi}), one finds that  the right hand side of Eq. (\ref{MZaxi}) vanishes at the boundary.  As a consequence, $M^Z$ also vanishes, leaving a single free parameter, $c$.

\subsubsection{Cylinders}

For a circular cylinder of radius $R_0$ one has $R'=0$ so that $Z=l$ and $Z'=\sin \alpha$. Now $C_\perp=0$, $C_\parallel=1/R_0$ and as a consequence $\kappa_g=\alpha'$, $\kappa_n = \cos^2 \alpha/R_0$ and $\tau_g= \sin 2 \alpha /(2 R_0)$.  In addition to the rotational symmetry about the axis, the geometry also possesses translational symmetry along this axis, so that the corresponding projection of $\mathbf{F}$, $F^Z = \mathbf{F} \cdot \hat{\bf k}$ is conserved. In general this projection reads
\begin{equation} \label{FZ}
F^Z = R C_\parallel \left(\sin \alpha F^{\bf T} + \cos \alpha F^{\bf L} \right) -R' \csc \alpha F^{\bf N}\,,
\end{equation}
which for the cylinder reduces to
\begin{equation} \label{FZcyl}
F^{Z} = \sin \alpha F^\mathbf{T} + \cos \alpha F^\mathbf{L}\,.
\end{equation}
The cylinder is flat so that ${\cal K}_G=0$; thus, in the absence of a potential,  $\lambda^\mathbf{L}=0$ according to Eq. (\ref{eq:lamL0}). Thus there is no tangential force along the curve. Taking into account these results, it is straightforward to confirm  that $F^Z{}' = \cos \alpha \, \varepsilon_\mathbf{L}$. $F^Z$ is conserved when the EL equation is satisfied.
\vskip1pc \noindent
It is now possible to exploit the two-dimensional subgroup of the Euclidean group unbroken by motion along the cylinder to identify a quadrature. In this case $M^Z$ given by expression (\ref{MZ}) simplifies to give
\begin{equation} \label{MZcyl}
M^Z/ R_0 = \cos \alpha \left( F^\mathbf{T} - {\cal H}_n /R_0 \right) - \sin\alpha \left( F^\mathbf{L} + {\cal T}_g /R_0 \right)\,,
\end{equation}
Both Eqs. (\ref{FZcyl}) and (\ref{MZcyl}) are of second order in derivatives of $\alpha$ (this dependence enters
through ${\cal H}'_g$ which involves $\kappa_g'$).  By taking an appropriate linear combination of $F^Z$ and $M^Z$, it is possible to eliminate the term involving this second derivative, $F^\mathbf{L}$. Specifically
 \begin{equation} \label{cylinder conserve}
F^Z \sin \alpha + M^Z/R_0\,  \cos \alpha = F^\mathbf{T}  - \kappa_n {\cal H}_n -\tau_g {\cal T}_g
= \kappa_g {\cal H}_g - {\cal H} - c\,,
\end{equation}
where the expression for $\kappa_n$ and $\tau_g$ on a cylinder have been used.
For an energy density quadratic in the geodesic curvature of the form ${\cal H}= 1/2 (\kappa_g-C_g)^2 + h(\kappa_n, \tau_g)$, Eq. (\ref{cylinder conserve})  provides a quadrature for $\alpha$:
\begin{equation} \label{2ndintcylconf}
\frac{1}{2}  (\alpha')^2 + U(\alpha) = \frac{1}{2}C^2_g + c\,,
\end{equation}
where the potential $U$ is given by
\begin{equation}
U(\alpha) = - h - F^Z \sin \alpha - M^Z/R_0 \, \cos \alpha\,,
\end{equation}
involving the two constants $F^Z$ and $M^Z$. The spontaneous curvature $C_g$ plays no role. For bound Euler Elastic curves, $h = \kappa^2_n/2 = \cos^4 \alpha/ (2 R^2_0)$.
\vskip1pc \noindent
The other natural linear combination of the conservation laws provides a useful expression for ${\cal H}'_g$, and with it a remarkably simple expression for the transmitted force, $\lambda^\mathbf{N}$. One has
\begin{equation} \label{Hgpcyl}
F^Z \cos \alpha - M^Z/R_0\, \sin \alpha = F^\mathbf{L} + \tau_g {\cal H}_n + \left(\frac{1}{R_0} -\kappa_n \right) \,{\cal T}_g = {\cal H}'_g + 2 \tau_g \, {\cal H}_n + \left(\frac{1}{R_0} - 2 \kappa_n \right) \,{\cal T}_g\,,
\end{equation}
Likewise, for bound Euler-elastic curves, Eq. (\ref{Hgpcyl})  reads
\begin{equation} \label{HgpcylEuEl}
\kappa'_g + 2 \kappa_n \, \tau_g - F^Z \, \cos \alpha + \frac{M^Z}{R_0} \, \sin \alpha=0\,.
\end{equation}
The magnitude of the force on the cylinder is given by
\begin{eqnarray} \label{lambdacyl}
\lambda^\mathbf{N} &=& {\cal H}''_n + \left[\sec \alpha (\sin \alpha {\cal T}_g)' + (\ln \cos \alpha)' {\cal H}_n -R_0 \sec^2 \alpha V' \right]' \nonumber\\
&+&\frac{\cos^2 \alpha}{R_0} \left(-\tan \alpha {\cal H}'_g + \alpha' {\cal H}_g + \frac{\cos 2 \alpha}{R_0} {\cal H}_n + \frac{\sin 2 \alpha}{R_0} {\cal T}_g -{\cal H}-V-c\right)\,,
\end{eqnarray}
For Euler elastica, using the quadrature (\ref{2ndintcylconf}) and the second order DE (\ref{HgpcylEuEl}) to eliminate the derivatives of $\alpha$ in (\ref{lambdacyl}), $\lambda^\mathbf{N}$  reduces to the form
\begin{equation} \label{lambdacylEuEl}
\lambda^\mathbf{N} = 2 \kappa^2_n \, \left(\frac{5}{R_0} -6 \kappa_n \right) + 3 \, F^Z \, \sin \alpha \left(\frac{2}{R_0} - 5 \kappa_n \right) + 5 \, \frac{M^Z}{R_0} \, \cos \alpha \left(\frac{2}{R_0} -3 \kappa_n\right) - 6 \frac{c}{R_0} \,\cos 2 \alpha \,.
\end{equation}
It depends only on $\alpha$. Confinement within cylinders will be treated in detail elsewhere.

\subsection{Glide rotations: Helicoids}\label{helicoid0}

A helicoid consists of an infinite double spiral staircase winding about a fixed axis \cite{Gray};  it can be  represented very simply in terms of a height function over the orthogonal plane; one half is described by $(r,\theta) \to h(r,\theta)= p \, \theta$, where $p$ is its pitch. Its other half  is described by $h(r,\theta)= p \, (\pi + \theta)$. The surface is smooth along $r=0$, which forms the boundary of each staircase.  It is straightforward to confirm that the helicoid is a minimal surface satisfying $\nabla^2 h=0$, where $\nabla^2$ is the Laplacian on the plane. Curves of constant $r$ form helices.
\vskip1pc \noindent
The symmetry of a helicoid is a glide rotation, i.e. the composition of an infinitesimal rotation by an angle $\delta \theta$ about the rotation axis,  with a translation by a height $\delta h = p \, \delta \theta$ along it, given by $\delta \mathbf{Y}= \delta \theta\, \hat{\bf k}\times \mathbf{Y}  + p \, \delta \theta \hat{\bf k}$. The relevant conserved quantity is then
\begin{equation}  \label{GZ}
G^Z = M^Z + p \,F^Z  = (\mathbf{M}+ p \,\mathbf{F})\cdot \mathbf{k}\,.
\end{equation}
One has
\begin{subequations}
\begin{eqnarray}
F^Z &=& \frac{1}{\sqrt{r^2 + p^2}} \left( p \left(\cos \beta F^\mathbf{T} + \sin \beta F^\mathbf{L}\right) +r F^\mathbf{N}\right)\,;\\
M^Z &=& \frac{1}{\sqrt{r^2 + p^2}} \left(\cos \beta (r^2 \, F^\mathbf{T}-p {\cal T}_g) + \sin \beta ( r^2 F^\mathbf{L} - p{\cal H}_n) - r (p F^\mathbf{N} +{\cal H}_g) \right)\,,
\end{eqnarray}
\end{subequations}
where $\beta$ is the angle that the tangent vector ${\bf T}$ makes with the helical direction $\hat{\bf e}_\theta$ (see \ref{Apphelicoid}).
Both projections involve all three components of $\mathbf{F}$. Therefore $G^Z$ is given by
\begin{eqnarray} \label{GZex}
G^Z &=& \frac{1}{\sqrt{r^2 + p^2}} \left( \cos \beta ((r^2+p^2) \, F^\mathbf{T}-p {\cal T}_g) + \sin \beta ( (r^2+p^2) F^\mathbf{L} - p{\cal H}_n) - r {\cal H}_g \right) \\
&=& \sqrt{r^2 + p^2} \left( \sin \beta \left( {\cal H}'_g + 2 \tau_g {\cal H}_n - 2 \kappa_n {\cal T}_g \right) +\left( \cos \beta \kappa_g -\frac{r}{r^2 + p^2} \right) {\cal H}_g - \cos \beta ({\cal H}+V+c ) \right)\,. \nonumber
\end{eqnarray}
The sum does not involve $F^\mathbf{N}$.  As shown in \ref{AppGZp}, its derivative is proportional to the EL derivative,  or $G^Z{}' = \sqrt{r^2 + p^2}\, \sin \beta \, \varepsilon_\mathbf{L}$, so that it is conserved in equilibrium and provides a first integral of the EL equation.
\vskip1pc \noindent
An alternative derivation for $M^Z$ and $G^Z$ within a Hamiltonian formulation is presented in \ref{AppHamForm}.

\section{Discussion}

The physics of linear polymers or protein complexes  bound  to fluid membranes is, in general,  enormously complicated and modeling their  behavior at a microscopic level involves a significant computational effort. However, just as essential aspects of the physics of fluid membranes on mesoscopic scales are captured by the geometrical degrees of freedom of the membrane surface,  elements of the interaction between the surface-bound protein and its substrate can be understood on such scales in terms of the geometrical  environment:  the linear structure  can be modeled as a curve on a surface with a potential energy that reflects its interaction with this environment.  The accommodation of such an environmental bias has never been examined in any systematic way, either by mathematicians or physicists, even though it is a natural question either from the point of view of dynamical systems or of differential geometry, and an obvious generalization of the study of geodesics.  Along the way, one is forced to 
think how curves best negotiate energetically costly obstacles on a surface consistent with its environmental biases. To address this problem, it is necessary to extend existing geometrical methods in the calculus of variations to accommodate elastic energies  capturing these biases.
\vskip1pc \noindent
We have shown how the conservation laws associated with any residual continuous Euclidean symmetry can be identified,  focusing not only on the obvious axially symmetric surfaces but also on surfaces--exemplified by a helicoid--with glide rotational symmetry. Teasing out the physics has involved mathematics of interest in its own right which warrants closer inspection.
\vskip1pc \noindent
It is beyond the scope of this paper to catalog the equilibrium states on even the simplest geometries. In future publications  we will examine the ground states of  physically interesting variants of the elastic energy (\ref{HBD})  on cylinders, catenoids and helicoids as well as some more exotic geometries. For spheres there is not a lot to say that was not already reported in \cite{GuvVaz}.\footnote{This is not the case for higher dimensional analogs of this problem, as demonstrated in Ref. \cite{Hyperspheres} where the confinement of spheres
within hyperspheres was examined.}  In our work thus far,  we do find time and time again that it is invaluable to first understand  how geodesics and curves of constant geodesic curvature behave on the surface.  Geodesics are not without interest themselves as the configurations of a taut string  negotiating the surface.  Curves of constant geodesic curvature, on the other hand,  may be interpreted as domain boundaries of a two-phase system living on the surface.  For helicoids and catenoids, 
there is a one to one mapping between such curves on one and those on the other:
this is because each turn of a helicoid is isometric to a catenoid\footnote{Animated nicely in its wikipedia entry \cite{HeltocatWiki}} so that the trajectories of constant geodesic curvature in one imply those in the other  \cite{DoCarmo, Gray},  even if the two are very different from a physical point of view. This distinction, as emphasized throughout this paper,  is captured by the normal forces. Of course,  from a mathematical point of view, the intrinsic aspect of this subject is classical;  what is  surprising is that it appears not to have been treated from a physically relevant point of view, that addresses the three-dimensionality of the problem.  Likewise, it is  useful to examine curves along which $\kappa_n$ or $\tau_g$ vanish or are constant. Of course, in an elastic system, none of these curves tend to appear as equilibrium states. In general, the constancy of any one of $\kappa_g$, $\kappa_n$ or $\tau_g$ will be incompatible with that of the other two. This frustration of access to the ground
state gives rise to interesting, and on occasion  counterintuitive, behavior.  In particular, geodesics tend to be incompatible with $\kappa_n=0$. Indeed, the later identity is only possible along asymptotic directions, which occur only if the local Gaussian curvature is non-positive.
\vskip1pc \noindent
An obvious omission in this paper is the response of the membrane to the elastic curve.  The surface generally will not play a passive role.  This is clearly true if the elastic curve forms the   boundary.  Whereas a line tension induced on the boundary of a fluid membrane will tend to heal the disruption caused by the edge, \cite{CGS, Tu0, Tu1} a boundary providing resistance to bending--or selecting some preferred curvature--can provide stable ends to an otherwise unstable membrane preventing its closure. The inner boundary of the recently discovered Terasaki ramps in the rough endoplasmic reticulum  are believed to be stabilized by  a mechanism of this kind  due to the condensation, along it,  of reticulons \cite{Terasaki}. Unfortunately, as anyone who has taken a moment to ponder these boundary conditions will have noted--even in the apparently simple scenario where the boundary energy is dominated by line tension--getting them right is a challenge.  The interaction between a fluid membrane and an 
elastic boundary will be taken up in a subsequent paper.
\vskip2pc \noindent
{\Large \bf Acknowledgements}
\vskip1pc \noindent
We would like to thank Markus Deserno, Martin M. M\"uller and Zachary McDargh for helpful discussions, as well as Nadir Kaplan for a useful pointer. This work was partially supported by CONACyT grant 180901. PVM acknowledges the support of CONACyT postdoctoral fellowship 205393.

\begin{appendix}

\setcounter{equation}{0}
\renewcommand{\thesection}{Appendix \Alph{section}}
\renewcommand{\thesubsection}{A. \arabic{subsection}}
\renewcommand{\theequation}{A. \arabic{equation}}

\section{Identity for \texorpdfstring{$\nabla_\mathbf{L} {\cal K}_G$}{nabla L KG} on a minimal surface}  \label{AppCM}

The identity (\ref{kKkkt}) between the normal derivative of $K_G$ and the behavior of the Darboux curvatures and torsion follows as a consequence of the Gauss-Codazzi-Mainardi equations. On a minimal surface the Gauss-Codazzi equation reads
\begin{equation} \label{GCK0}
{\cal K}_G = -1/2 K_{ab} K^{ab} = -(\kappa_n^2+\tau_g^2)\,.
\end{equation}
Taking the directional derivative along the normal to the curve $\mathbf{L}$ and using the Codazzi-Mainardi equations $\nabla_a K_{bc} = \nabla_b K_{ac}$, one has
\begin{eqnarray}
\nabla_\mathbf{L} {\cal K}_G &=& - K^{ab} l^c \nabla_c K_{ab} = - K^{ab} l^c \nabla_a K_{bc} \nonumber\\
&=& -K^{ab} \left[ \nabla_a (l^c K_{bc})- K_{bc} \nabla_a l^c\right]\,.
\label{B2}\end{eqnarray}
On  a minimal surface the extrinsic curvature tensor can be expanded with respect to the tangential Darboux basis vectors $\{\mathbf{T},\mathbf{L}\}$ as $K_{ab} = (t_a t_b - l_a l_b)\kappa_n - (t_a l_b +t_b l_a ) \tau_g$, where $t^a$ and $l^a$ are their projections onto the basis adapted to the parametrization of the surface. It follows that $l^c K_{bc}= - l_b \kappa_n -t_b \tau_g$. Using these expressions along with the definition of the geodesic curvature $\kappa_g = l^a t^b \nabla_b t_a$, one finds for each of the two term on the second line in Eq. (\ref{B2}),
\begin{subequations}
\begin{eqnarray}
-K^{ab} \nabla_a (l^c K_{bc}) &=& \kappa_n \tau'_g - \kappa'_n \tau_g + \kappa_g {\cal K}_G + 1/2 \nabla_\mathbf{L} {\cal K}_G \,; \\
K^{ab} K_{bc} \nabla_a l^c &=& \kappa_g {\cal K}_G\,.
\end{eqnarray}
\end{subequations}
Summing terms and simplifying, Eq. (\ref{kKkkt}) is identified.

\setcounter{equation}{0}
\renewcommand{\thesection}{Appendix \Alph{section}}
\renewcommand{\thesubsection}{B. \arabic{subsection}}
\renewcommand{\theequation}{B. \arabic{equation}}

\section{Curves on axially symmetric surfaces} \label{axisymm}

\begin{figure}[htb]
\begin{center}
\includegraphics[scale=0.75]{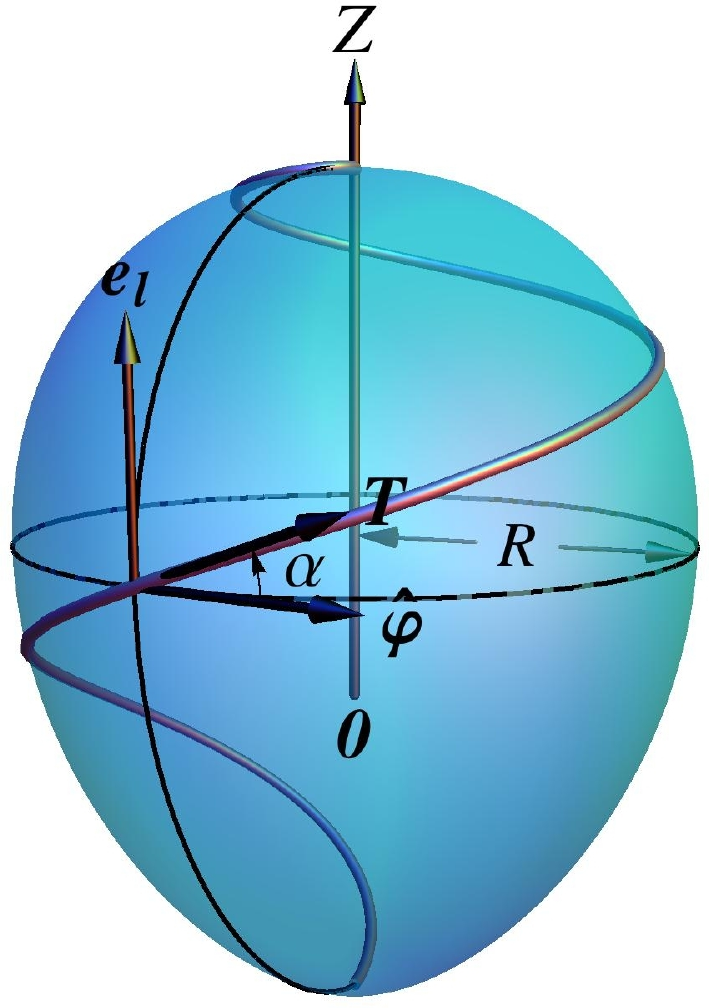}
\end{center}
\caption{\small{Curve on an axisymmetric surface. The tangent basis adapted to the surface is ${\bf e}_l$ and ${\bf e}_\varphi=R \hat{\bm \varphi}$. The tangent vector ${\bf T}$ makes an angle $\alpha$ with the azimuthal direction $\hat{\bm \varphi}$.}}
\label{Fig2}
\end{figure}

An axially-symmetric surface is described by its embedding
\begin{equation}
\Sigma:(l,\varphi) \rightarrow \mathbf{X}(l,\phi)= (R(l)\cos \varphi, R(l)\sin\varphi, Z(l))
\end{equation}
into three-dimensional space, where $l$ is arc length along the meridian and $\varphi$ is the polar angle along the parallel. $R(l)$ is the polar radius, and $Z(l)$ the corresponding height, see Fig. \ref{Fig2}. The two tangent vectors adapted to this parametrization will be denoted $\mathbf{e}_l$ and $\mathbf{e}_\phi$. They also form the principal directions on the surface with corresponding curvatures $C_\perp$ along the meridian, and $C_\|$ along the parallel.
A curve parametrized by arc-length $s$ on such a surface is described by the embedding $s\to (l(s),\varphi(s))$, or $\Gamma:s \rightarrow \mathbf {Y}(s) = R(s) \hat{\bm \rho}(s) + Z(s)\hat{\bf k}$.\footnote {The shorthand $R(s)$ for  $R(l(s))$, and similarly for $Z$, is understood.}, where $\hat{\bm \rho}(s) = (\cos\varphi(s), \sin\varphi(s),0)$ is the unit vector along the polar radial direction. The corresponding unit vector in the azimuthal direction is $\hat{\bm \varphi}(s) = (-\sin\varphi(s), \cos\varphi(s), 0)$. On our surface $R$ and $Z$ will depend on $s$ through $l$. It is straightforward to expand the tangent vector along the curve $\mathbf{T} =\mathbf{Y'}$ (where the prime denotes differentiation with respect to arc-length, $\partial_s$)  as well as its Darboux
counterpart $\mathbf{L} = \mathbf{N} \times \mathbf{T}$,  with respect to the the adapted basis vectors:
\begin{equation}
\mathbf{T} = \cos \alpha \, \hat{\bm \varphi}  + \sin \alpha \, \mathbf{e}_l  \,, \qquad \mathbf{L} = -  \sin \alpha \, \hat{\bm \varphi} + \cos \alpha \, \mathbf{e}_l\,.
\end{equation}
The unit vector tangent to the curve is now characterized by the angle $\alpha(s)$ that it makes with the parallel direction $\hat{\bm \varphi}$; parametrization by arc-length implies ($l'{}^2= R'{}^2 + Z'{}^2 $)
\begin{equation}
R'{}^2  + Z'{}^2 + R^2 \varphi'{}^2 =1 \,,
\end{equation}
so that Eqs. (\ref{lprime}) and (\ref{phiprime}) follow. The principal curvatures on the curve are given by
\begin{equation} \label{CparCperp}
 C_\parallel = \csc \alpha \, \frac{Z'}{R} \,, \quad C_\perp = \csc^3 \alpha \left(R' Z'' - Z' R''\right)\,.
\end{equation}
The two surface unit tangent vectors change along the curve as
\begin{equation} \label{elephiald}
\hat{\bm \varphi}' = -\cot \alpha R'/R \, \mathbf{e}_l -\cos \alpha C_\parallel \, \mathbf{N} \,, \quad \mathbf{e}'_l = \cot \alpha R'/R \, \hat{\bm \varphi} -\sin \alpha C_\perp \, \mathbf{N} \,.
\end{equation}
Using these expressions one can decompose the acceleration with respect to the Darboux basis as:
\begin{equation}
\mathbf{T}' = ({\alpha}' - \frac{ R'}{R} \, \cot{\alpha}) \, \mathbf{L} - (\sin^2 \alpha \, C_\perp + \cos^2 \alpha \,
C_\parallel) \, \mathbf{N}\,.
\end{equation}
This decomposition identifies the geodesic and normal curvatures. Firstly, one has that the geodesic curvature of $\Gamma$: $\kappa_g = \mathbf{T}' \cdot \mathbf{L} $  is given by
\begin{equation} \label{axikappag}
\kappa_g = - \frac{(R\,\cos \alpha)'}{R\sin \alpha}\,.
\end{equation}
If $\alpha\ne 0$, geodesic curves satisfy the remarkably simple Clairaut's relationship \cite{DoCarmo, Arnold}
\begin{equation} \label{Clairaut}
\cos \alpha = {\cal C} /R \,,
\end{equation}
where ${\cal C}$ is constant with dimensions of distance. In particular,  meridians with constant $\alpha=\pi/2$ (and extremal parallels with $d R/dl=0$, when they exist) are also geodesic. Under the change of chirality, $\alpha\to \pi-\alpha$, $\kappa_g$ changes sign: $\kappa_g\to -\kappa_g$.
\vskip1pc \noindent
The corresponding normal curvature  $\kappa_n = - \mathbf{T}' \cdot \mathbf{N}$  is given in terms of the angle $\alpha$  and the principal curvatures, $C_\perp$ and $C_\|$, by Euler's formula,
\begin{equation} \label{axikappan}
\kappa_n = \sin^2 \alpha \, C_\perp + \cos^2 \alpha \,C_\parallel\,.
\end{equation}
If the Gaussian curvature at a point is negative, so that ${\cal K}_G=C_\perp C_\| <0$, $\kappa_n$ will vanish along the tangential directions given by $\tan \alpha = \sqrt{-C_\parallel/C_\perp}$. On a minimal surface, $\alpha=\pi/4$ along such curves.
The geodesic torsion $\tau_g = \mathbf{L}' \cdot \mathbf{N}$ assumes the form
\begin{equation} \label{axitaug}
 \tau_g = \sin \alpha \cos \alpha (C_\parallel - C_\perp)\,.
\end{equation}
It vanishes when the tangent is aligned along a principal direction. Under the change of chirality $\alpha\to \pi-\alpha$: $\tau_g\to -\tau_g$, whereas $\kappa_n$ is unchanged.

\setcounter{equation}{0}
\renewcommand{\thesection}{Appendix \Alph{section}}
\renewcommand{\thesubsection}{C. \arabic{subsection}}
\renewcommand{\theequation}{C. \arabic{equation}}

\section{Calculation of \texorpdfstring{$M^Z{'}$}{MZ'}} \label{AppMZp}

By either projecting Eq. (\ref{Mp}) onto $\hat{\bf k}$, or differentiating $M^Z$ given by Eq. (\ref{MZ}), one obtains
\begin{equation}
 M^Z{}' = R \sin \alpha \left( - \varepsilon_\mathbf{L} + \lambda^\mathbf{L} +C_\parallel (\Lambda^\mathbf{L} -\cot \alpha \Lambda^\mathbf{T})\right)\,,
\end{equation}
We need to show that the second two terms sum to zero.
We do this by showing that
\begin{equation}\label{Idef}
 I:= \lambda^\mathbf{L} + C_\parallel (\Lambda^\mathbf{L} -\cot \alpha \Lambda^\mathbf{T}) =  -\cot \alpha V' +\nabla_\mathbf{L} V
\end{equation}
vanishes along curves on axisymmetric surfaces because  $\nabla_\mathbf{L} V = \cot \alpha V'$
if $V$ depends only on the meridian arc length $l$. Thus one has that $M^Z{}'= -R \sin \alpha \varepsilon_\mathbf{L}$.
\vskip1pc \noindent
To establish Eq. (\ref{Idef}), we use the identity Eq. (\ref{eq:lampar}) for $\lambda^\mathbf{L}$ along with Eqs. (\ref{axikappan}) and (\ref{axitaug})  defining $\kappa_n$ and $\tau_g$ along curves on axisymmetric surfaces,  to obtain
\begin{equation}
I =  -\cot \alpha \frac{C_\parallel}{\kappa_n} \left( \kappa_n \Lambda^\mathbf{T} - \tau_g \Lambda^\mathbf{L}\right) + \frac{\tau_g}{\kappa_n} V' +\nabla_\mathbf{L} V\,.
\end{equation}
Using the identity Eq. (\ref{LamTLamL}) for
$\kappa_n \Lambda^\mathbf{T} - \tau_g \Lambda^\mathbf{L}$, as well as the identity $\tau_g = \cot \alpha (C_\parallel -\kappa_n)$,
Eq. (\ref{Idef}) follows.

\setcounter{equation}{0}
\renewcommand{\thesection}{Appendix \Alph{section}}
\renewcommand{\thesubsection}{D. \arabic{subsection}}
\renewcommand{\theequation}{D. \arabic{equation}}

\section{Helicoids} \label{Apphelicoid}

The embedding functions describing a helicoid parametrized by a polar chart on the plane $(r, \theta)$ centered on the rotation axis is given by
${\bf X}_H(r, \theta) = r \hat{\bf r} + p \theta \hat{\bf k}$, where $\hat{\bf r}=(\cos \theta, \sin \theta,0)$; $-\infty <r < \infty$.\footnote{Extending the range of $r$ permits one to treat the two
spiral staircases as a single double spiral staircase.}  One revolution is described by $-\pi \le \theta \le \pi$; the constant $p$ characterizes the pitch; if positive the helicoid is right handed; if negative, it is left handed. As $p\rightarrow 0$ the helicoid degenerates into the plane.
\vskip1pc \noindent
The tangent vectors adapted to the helicoid in this parametrization are given by ${\bf e}_r = \hat{\bf r}$ and ${\bf e}_\theta = r \hat{\bm \theta} + p \hat{\bf k}$, $\hat{\bm \theta} = (-\sin \theta, \cos \theta,0)$. Whereas ${\bf e}_r$ is a unit vector,  ${\bf e}_\theta $ is not; $\hat{\bf e}_\theta = {\bf e}_\theta/(r^2 + p^2 )^{1/2}$ is normalized.  The line element is $ds^2 = dr^2 + (r^2 +p^2) d \theta^2$.
\vskip1pc \noindent
The unit normal vector is $\mathbf{n} = (-p \,\hat{\bm \theta} + r\, \hat{\bf k})/ (r^2 + p^2)^{1/2}$.
\vskip1pc \noindent
The corresponding curvatures are $C_{\pm} = \pm p/(r^2 + p^2)$, so that the Gaussian curvature is given by
\begin{equation} \label{KGhel}
 {\cal K}_G= -p^2/(r^2 + p^2)^2\,,
\end{equation}
decaying rapidly with distance $r$ along the rulings.

\subsection{Curves on the helicoid}

\begin{figure}[htb]
\begin{center}
\includegraphics[scale=0.75]{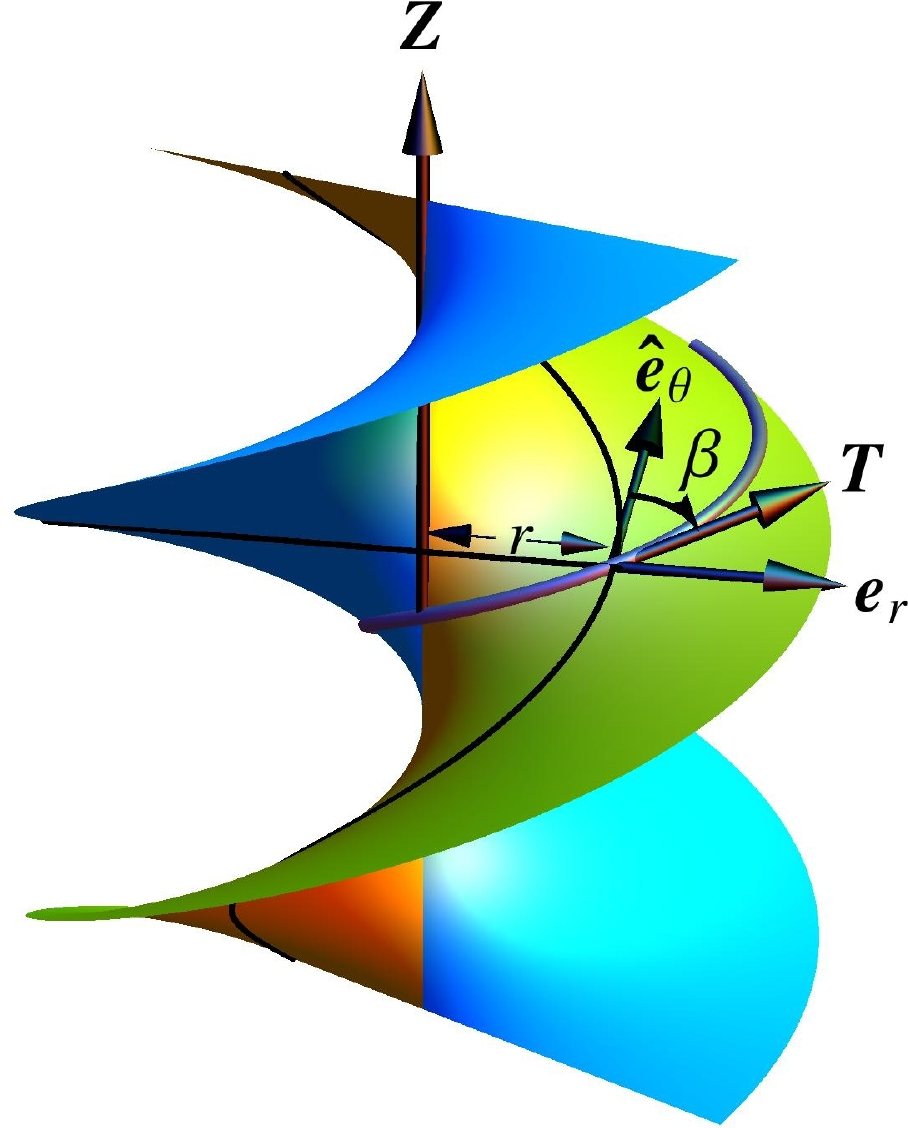}
\end{center}
\caption{\small{Curve on the helicoid. The tangent basis adapted to the surface is ${\bf e}_r$ and ${\bf e}_\theta$. The tangent vector ${\bf T}$ makes an angle $\beta$ with the azimuthal direction $\hat{\bf e}_\theta$.}}
\label{Fig3}
\end{figure}

Consider a curve on the Helicoid, $s\mapsto  = r(s) \hat{\bf r}(s) + p \, \theta(s) \hat{\bf k}$, parametrized by arc-length. Its unit tangent vector can be expanded with respect to the adapted basis vectors, $\mathbf{T}= r' \mathbf{e}_r + \theta' \mathbf{e}_\theta$, which can be written as $\mathbf{T}= t^a \mathbf{e}_a$, with components $t^a=(r', \theta')$. Arc-length parametrization implies the normalization $r'{}^2 + (r^2 + p^2) \, \theta'{}^2 = 1$. The surface normal to the curve is $\mathbf{L} = - \theta' (r^2 + p^2)^{1/2} \, {\bf e}_r + r' \,\hat{\bf e}_\theta$. Thus $l^a = (-\theta' \, (r^2 + p^2)^{1/2}, r'/(r^2+p^2)^{1/2})$. The curve can be characterized by
$\beta(s)$, the angle that ${\bf T}$ makes with the direction $\hat{\bf e}_\theta$, see Fig \ref{Fig3}.
In this parametrization the Darboux tangent basis, $\mathbf{T}$ and $\mathbf{L}$, is expressed as
\begin{equation}
\mathbf{T} = \cos \beta \, \hat{\bf e}_\theta + \sin \beta \, {\bf e}_r \,, \qquad \mathbf{L} = \sin \beta \, \hat{\bf e}_\theta - \cos \beta \,{\bf e}_r\,,
\end{equation}
from which one identifies the relations $\sin \beta = r'$ and $\cos \beta = (r^2 + p^2)^{1/2} \theta'$.
\vskip1pc \noindent
The geodesic curvature of the curve, $\kappa_g = \mathbf{T}' \cdot\mathbf{L}$  on the helicoid is
\begin{equation}\kappa_g = - \beta' + \frac{r \, \cos\beta}{r^2 + p^2} \,.
\label{kghel}
\end{equation}
It can also be deduced from its counterpart on the catenoid, defined by Eq. (\ref{axikappag}),  using the isometry between the two geometries, with the replacements $R = (r^2+p^2)^{1/2}$ and $\alpha = - \beta$.\footnote{$\alpha$ is measured with respect to $\hat{\bf e}_{\varphi}$ in an anticlockwise sense, whereas $\beta$ is measured clockwise with respect to $\hat{\bf e}_\theta$.} Geodesics, as before, satisfy a Clairaut type relation
\begin{equation}
(r^2 + p^2 ) \cos^2 \beta = {\cal C}^2\,.
\end{equation}
The normal curvature $\kappa_n = K_{ab} t^a t^b  $ is given by
\begin{equation} \label{kappanhel}
\kappa_n= \frac{p\, \sin 2 \beta}{ r^2+ p^2} \,.
\end{equation}
Thus the asymptotic directions coincide with the parameter curves with $\beta=0, \pi/2$. In particular, helices are  asymptotic.
The geodesic torsion $\tau_g = - K_{ab} t^a l^b$ is given by
\begin{equation} \label{taughel}
\tau_g = \frac{p \, \cos 2 \beta}{ r^2+ p^2} \,.
\end{equation}

\setcounter{equation}{0}
\renewcommand{\thesection}{Appendix \Alph{section}}
\renewcommand{\thesubsection}{E. \arabic{subsection}}
\renewcommand{\theequation}{E. \arabic{equation}}

\section{Calculation of \texorpdfstring{$G^Z{'}$}{GZ'}} \label{AppGZp}

Differentiating $G^Z$ given by Eq. (\ref{GZ}), one gets
\begin{equation}
G^Z{}' = \sqrt{r^2 + p^2} \sin \beta \left( \varepsilon_\mathbf{L} - \lambda^\mathbf{L} - \frac{p}{r^2 + p^2} (\Lambda^\mathbf{T} - \cot \beta \Lambda^\mathbf{L})\right)\,,
\end{equation}
Using Eq. (\ref{LamTLamL}) along with expression (\ref{eq:lampar}) for $\lambda^\mathbf{L}$ one gets
\begin{eqnarray}
 \lambda^\mathbf{L} + \frac{p}{r^2 + p^2} (\Lambda^\mathbf{T} -\cot \beta \Lambda^\mathbf{L}) &=&  \left( \frac{p}{r^2 + p^2} \left( \frac{\tau_g}{\kappa_n} -\cot \beta \right) -\frac{{\cal K}_G}{\kappa_n} \right) \Lambda^\mathbf{L} \nonumber \\
&+& \left(\frac{p}{r^2 + p^2} + \tau_g \right) \frac{V'}{\kappa_n} +\nabla_\mathbf{L} V\,.
\end{eqnarray}
From Eqs. (\ref{KGhel}), (\ref{kappanhel}) and (\ref{taughel}) for ${\cal K}_g$, $\kappa_n$ and $\tau_g$ of the helicoid, one finds that
\begin{eqnarray}
\frac{p}{r^2 + p^2} \left( \frac{\tau_g}{\kappa_n} -\cot \beta \right) -\frac{{\cal K}_G}{\kappa_n} &=& \frac{p}{r^2 + p^2} \left( \cot 2 \beta - \cot \beta + \csc 2 \beta \right)=0\,,\\
\frac{1}{\kappa_n} \left( \frac{p}{r^2 + p^2} +\tau_g \right) &=& \cot \beta \,.
\end{eqnarray}
Thus
\begin{equation}
 \lambda^\mathbf{L} + \frac{p}{r^2 + p^2} (\Lambda^\mathbf{T} -\cot \beta \Lambda^\mathbf{L}) =  \cot \beta V' +\nabla_\mathbf{L} V\,,
\end{equation}
which vanishes on account that for a curve on the helicoid $V$ depends only on the radial coordinate $r$, so that $\nabla_\mathbf{L} V = -\cot \beta \, V'$. Thus one has that $G^Z{}'= \sqrt{r^2 + p^2} \, \sin \beta \, \varepsilon_\mathbf{L}$.

\setcounter{equation}{0}
\renewcommand{\thesection}{Appendix \Alph{section}}
\renewcommand{\thesubsection}{F. \arabic{subsection}}
\renewcommand{\theequation}{F. \arabic{equation}}

\section{Hamiltonian framework adapted to symmetry} \label{AppHamForm}

\subsection{Elastic curves on surfaces with axial symmetry}

Consider an energy of the general form (\ref{Hcurve}),  with $V=0$ to avoid clutter.
The Lagrangian density  along a curve on an axially symmetric surface is given by
\begin{equation}
{\cal H}_C = {\cal H} [\alpha, \alpha', l,l',\varphi', \lambda_l, \lambda_\varphi] + \, \lambda_l \left(l' - \sin \alpha \right) + \, \lambda_\phi \left(\varphi'-\cos \alpha /R(l) \right)\,,
\end{equation}
where $l$ is arc-length along a meridian, $\varphi$ is the azimuthal angle, $R$ the polar radius, and $\alpha$ the angle the tangent makes with the polar radial direction.  For details see \ref{axisymm}.
The two constraints capture the parametrization by arc-length. The dependence of ${\cal H}$ on the generalized coordinates and their derivatives is not arbitrary, but enters through the curvatures and the torsion,  given for an axisymmetric surface by (dot is $\partial/\partial l$)
\begin{subequations}
\begin{eqnarray}
\kappa_g (\alpha, \alpha', l) &=& \alpha' - \cos \alpha\, \left(\ln R(l)\right)\dot{}\,,\\
\kappa_n(\alpha,l) &=& \sin^2 \alpha \, C_\perp(l) +\cos^2 \alpha \, C_\parallel(l)\,\\
\tau_g (\alpha,l) &=& 1/2 \sin 2 \alpha \, (C_\parallel(l) - C_\perp(l))\,.
\end{eqnarray}
\end{subequations}
where $C_\perp(l) = \ddot{Z} \dot{R} - \ddot{R} \dot{Z}$ and $C_\parallel (l) = \dot{Z}/R$.
The momentum densities conjugate to the three generalized coordinates, $(\alpha, l, \varphi)$, are now given by
\begin{equation} \label{PalphaHg}
 P_\alpha = {\cal H}_g\,, \qquad P_l =  \, \lambda_l \,, \qquad P_\varphi = \lambda_\varphi\,.
\end{equation}
${\cal H}$ does not depend explicitly on arc-length $s$, thus the Hamiltonian density ${\mathscr H}= \alpha' P_\alpha + l' P_l + \phi' P_\phi -{\cal H}_C$ is constant. Using the relations $l'= \sin \alpha$ and $\varphi' = \cos \alpha/R$, one finds it is given explicitly by
\begin{equation}\label{Hamaxi}
\mathscr{H} = \left(\kappa_g + \cot \alpha\, (\ln R)' \, \right) \, P_\alpha +  \sin\alpha \,P_l + \cos \alpha \, \frac{P_\phi}{R} -  {\cal H} \,.
\end{equation}
The Hamilton equations for the conjugate momenta are
\begin{subequations}
\begin{eqnarray}
R \left(\frac{P_\alpha}{R}\right)' &=& - 2 \tau_g {\cal H}_n - \left(K - 2 \kappa_n \right) \, {\cal T}_g- \cos \alpha P_l + \sin \alpha \frac{P_\varphi}{R}\,, \label{Palphap}\\
\sin \alpha P'_l &=& \left(\kappa'_g -\frac{(R\, \alpha')'}{R} \right) P_\alpha  + \left( \kappa'_n + 2 \alpha' \tau_g \right){\cal H}_n \nonumber \\
&&+ \left( \tau'_g + \alpha' (K - 2 \kappa_n) \right) {\cal T}_g + R' \cos \alpha \frac{P_\varphi}{R^2}\,, \label{Plp}\\
P'_\varphi&=&0 \label{Pvarphip}\,.
\end{eqnarray}
\end{subequations}
$P_\varphi$ is constant on account of the axial symmetry. $P_l$ is not in general constant and can be eliminated by taking an appropriate combination of Eqs. (\ref{Hamaxi}) and (\ref{Palphap}). Specifically, taking the combination $\sin \alpha P'_\alpha + \cos \alpha {\mathscr H}$ it is possible to express the constant $P_\varphi$ in terms of the remaining canonical variables:
\begin{equation} \label{Pvarphiaxi}
P_\varphi = R \sin \alpha \left( P'_\alpha + 2 \tau_g {\cal H}_n +\left( K- 2 \kappa_n \right)\, {\cal T}_g  \right) - (R \sin \alpha)' P_\alpha + R \cos \alpha \left( {\cal H} + \mathscr{H}\right)\,.
\end{equation}
Using the definitions of the remaining canonical momenta in  (\ref{PalphaHg}), along with the identifications $P_\phi = -M^Z$ and $\mathscr{H} = c$,  the ``first''  integral for a curve, given by Eq. (\ref{MZaxi}), is reproduced.
Thus, the ``Hamiltonian'' is identified as the constant of integration associated with fixed arc-length and the component of the torque along the axis of symmetry is minus the momentum conjugate to the coordinate along that axis.
\vskip1pc \noindent
For  the energy,  quadratic and symmetric in the Darboux curvatures, Eq. (\ref{Pvarphiaxi}) reads
\begin{eqnarray}
P_\varphi &=& R \sin \alpha \kappa'_g +\left(R/2 \cos \alpha (\kappa_g -C_g)- (R \sin \alpha)'\right) (\kappa_g - C_g) \nonumber \\
&&+ \mu\,R \left(\kappa_n -C_n \right) \left( 2 \sin \alpha \tau_g + \cos \alpha /2 (\kappa_n -C_n)\right) \nonumber \\
&& + \nu \, R \left(\tau_g -C_0 \right) \left( \sin \alpha (K - 2 \kappa_n) + \cos \alpha /2 (\tau_g - C_0)\right) +R c \cos \alpha\,.
\end{eqnarray}
To reconstruct the curve, one needs to solve the differential equation for $P_\alpha$
\begin{equation}
P'_\alpha = \left(\ln (R \sin \alpha)\right)' P_\alpha - 2 \tau_g {\cal H}_n -\left( K - 2 \kappa_n \right) \, {\cal T}_g + \csc \alpha \frac{P_\varphi}{R} - \cot \alpha ({\cal H} + \mathscr{H})\,,
\end{equation}
subject to the relation (\ref{PalphaHg}) which provides a relation between $\alpha'$ and $P_\alpha$. The conjugate moment $P_l$ can be determined from the combination $\sin \alpha  {\mathscr H} - \cos \alpha P'_\alpha$, obtaining
\begin{equation}
P_l = - \cos \alpha  \left( P_\alpha' + 2  \tau_g {\cal H}_n + \left(K - 2 \kappa_n \right) \, {\cal T}_g \right) - \sin \alpha \left( \kappa_g \, P_\alpha - {\cal H} - \mathscr{H}\right) \,.
\end{equation}
Using expressions (\ref{FTNL}) this can be recast as
\begin{equation}
-P_l = \sin \alpha F^{\bf T} + \cos \alpha F^{\bf L} + \left(\cos \alpha (K-\kappa_n) -\sin \alpha  \, \tau_g \right) {\cal T}_g - \sin \alpha \, C_\perp\,,
\end{equation}
so the component of ${\bf F}$ along the symmetry axis, given by (\ref{FZ}), can be written as
\begin{equation} \label{FZham}
F^Z = -R C_\parallel \left(P_l + \left(\cos \alpha (K-\kappa_n) -\sin \alpha  \, \tau_g \right) {\cal T}_g - \sin \alpha \, C_\perp \right) -R' \csc \alpha F^{\bf N}\,.
\end{equation}
Note the shortcoming of this approach: it does not provide the force.

\subsubsection{Cylindrical constraint} \label{Appcylhamform}

Here $R=R_0$, a constant, $C_\perp=0$, and $C_\|=1/R_0$. From Eq. (\ref{Plp}) follows that $P_l$ is constant (Eq. (\ref{FZham}) identifies it as minus the $F^{Z}$). For an energy quadratic in the geodesic curvature of the form ${\cal H} = 1/2 (\kappa_g -C_g)^2 + h(\kappa_n,\tau_g)$, Eq. (\ref{Hamaxi}) reduces to a genuine quadrature,
\begin{equation}
{\mathscr H} = \frac {1}{2} \alpha'{}^2 + V(\alpha) \,,
\end{equation}
where
\begin{equation}
V(\alpha) = - h + P_l \sin \alpha + \frac{P_\phi}{R_0} \cos \alpha -
\frac{1}{2} \, C_g^2\,.
\end{equation}
Taking into account the identifications $\mathscr{H}=c$, $P_l=-F^Z$ and $P_\varphi=-M^Z$, this reproduces the ``second'' integral Eq. (\ref{2ndintcylconf}).

\subsubsection{Catenoidal constraint} \label{Appcathamform}

Here $R(Z)= R_0\cosh (Z/R_0)$ or $R(l)= (l^2 + R_0^2)^{1/2}$ and $Z(l)= R_0 \arcsinh (l/R_0)$, therefore $C_\|= R_0/R^2=-C_\perp$ and $R' = \sin \alpha \, \tanh Z/R_0$, thus $\kappa_g = \alpha' - 1/R_0 \cos \alpha \, \sech Z/R_0 \, \tanh Z/R_0$, $\kappa_n = R_0 /R^2 \, \cos 2 \alpha$ and $\tau_g = R_0/R^2 \sin 2 \alpha$. For the quadratic and symmetric energy in the Darboux curvatures, Eq. (\ref{HBD}), the "first integral", Eq. (\ref{Pvarphiaxi}) reads
\begin{eqnarray} \label{PvarphiCat}
P_\varphi &=& R \sin \alpha \alpha'' - R/2 \cos \alpha \, \alpha'{}^2 + \sin^2 \alpha \cos \alpha/R (\tanh^2 Z/R_0 -\sech^2\,Z/R_0) \nonumber \\
&& + \left(\cos \alpha/R \, \tanh Z/R_0 + C_g \right) \left(R/2 \cos \alpha \left(\cos \alpha/R \, \tanh Z/R_0 + C_g \right) + R' \sin \alpha \right) \nonumber \\
&& + \mu\,R \left(\kappa_n -C_n \right) \left( 2 \sin \alpha \tau_g + \cos \alpha /2 (\kappa_n -C_n)\right) \nonumber \\
&& + \nu \, R \left(\tau_g -C_0 \right) \left( -2 \sin \alpha \kappa_n + \cos \alpha /2 (\tau_g - C_0)\right) +R c \cos \alpha\,.
\end{eqnarray}

\subsection{Helicoidal constraint}

The effective energy is
\begin{eqnarray}
{\cal H}_C = {\cal H}[\beta, \beta', r, r', \theta', \lambda_l, \lambda_\theta] + \lambda_r \left(r' - \sin \beta \right) +  \lambda_\theta \left(\theta' - \cos \beta/\sqrt{r^2 +p^2} \right)\,.
\end{eqnarray}
The momentum densities conjugate to the three generalized coordinates, $(\alpha, l, \varphi)$, are given by
\begin{equation} \label{PbetaHg}
 P_\beta = -{\cal H}_g\,, \qquad P_r =  \, \lambda_r \,, \qquad P_\theta = \lambda_\theta\,.
\end{equation}
${\cal H}$ does not depend explicitly on arc-length $s$, thus the Hamiltonian density ${\mathscr H}= \beta' P_\beta + r' P_r + \theta' P_\theta - {\cal H}_C$ is constant. Using the relations $r'= \sin \beta$ and $\theta' = \cos \beta/\sqrt{r^2 +p^2}$, one finds it is given explicitly by
\begin{equation}\label{Hamaxihel}
\mathscr{H} = P_\beta \, \left(-\kappa_g + r \cos \beta\,/ \sqrt{r^2 + p^2} \, \right)  + P_r \, \sin\beta + P_\theta \, \cos \beta/\sqrt{r^2+p^2} -  {\cal H} \,.
\end{equation}
The Hamilton equations for the conjugate momenta are
\begin{subequations}
\begin{eqnarray}
\sqrt{r^2 + p^2} \left(\frac{P_\beta}{\sqrt{r^2 + p^2}}\right)' &=& 2 \left(\tau_g {\cal H}_n - \kappa_n \, {\cal T}_g \right) - P_r \, \cos \beta + P_\varphi \, \sin \beta/\sqrt{r^2 + p^2}\,, \label{Pbetap}\\
(r^2+p^2)\,P'_r &=& P_\beta \cos \beta \frac{r^2-p^2}{r^2+p^2} + P_\theta \frac{r \cos \beta}{\sqrt{r^2+p^2}} -2 \, \frac{r}{p} \, \left( \kappa_n {\cal H}_n + \tau_g {\cal T}_g \right)\,, \label{Prp}\\
P'_\theta&=&0 \label{Pvarthetap}\,.
\end{eqnarray}
\end{subequations}
$P_\theta$ is constant on account of the axial symmetry. $P_r$ is not in general constant but can be eliminated by taking an appropriate combination of Eqs. (\ref{Hamaxihel}) and (\ref{Pbetap}). Specifically from the combination $\sin \beta P'_\beta + \cos \beta {\mathscr H}$ one can solve for the constant $P_\theta$ obtaining
\begin{equation} \label{Pvarthetaaxi}
-P_\theta = \sqrt{r^2 + p^2} \left[ \sin \beta \left(- P'_\beta + 2 \tau_g {\cal H}_n - 2 \kappa_n \, {\cal T}_g \right) - \cos \beta \left( {\cal H} + \mathscr{H}\right) \right] + P_\beta \left(\sqrt{r^2 + p^2} \, \sin \beta \right)'\,.
\end{equation}
Taking into account the relation (\ref{PbetaHg}), along with the identifications $P_\theta = -M^Z$ and $\mathscr{H} = c$, the ``first'' integral for an elastic curve on a helicoid, given by Eq. (\ref{GZex}), is reproduced.
Thus, the ``Hamiltonian'' is again identified as the constant of integration associated with fixed arc-length and the component of the torque along the axis of symmetry is minus the momentum conjugate to the coordinate along that axis.
\vskip1pc \noindent
To reconstruct the curve, one has to solve the ODE for $P_\beta$
\begin{equation}
P'_\beta = P_\beta \left( \ln (\sqrt{r^2 + p^2} \, \sin \beta)\right)' + 2 \left(\tau_g \, {\cal H}_n - \kappa_n {\cal T}_g\right) + \csc \beta \frac{P_\theta}{\sqrt{r^2 + p^2}} - \cot \beta ({\cal H} + \mathscr{H})\,;
\end{equation}
the identification in Eq. (\ref{PbetaHg}) provides a relation between $\beta'$ and $P_\beta$. The conjugate moment $P_r$ can be determined from the combination $\sin \beta  {\mathscr H} - \cos \beta P'_\beta$, obtaining
\begin{equation}
P_r = \cos \beta \left(- P_\beta' + 2 \left(\tau_g {\cal H}_n - \kappa_n  {\cal T}_g\right) \right)+ \sin \beta \left( \kappa_g \, P_\beta + {\cal H} + \mathscr{H}\right) \,.
\end{equation}

\end{appendix}

\end{document}